\documentclass{jfm}
\usepackage{graphicx}
\usepackage{epstopdf, epsfig}

\newcommand{\ts}{\textsuperscript}
\usepackage{xcolor}
\usepackage{amsmath} %use integral symbol (iiit_)
\usepackage[section]{placeins}
\shorttitle{Third-order law in MHD turbulence}
\shortauthor{Bin Jiang, Cheng Li, Yan Yang and others }

\title{Energy transfer and third-order law in forced anisotropic MHD turbulence with hyperviscosity} % BJ: a bit long

\author{Bin Jiang\aff{1,2}, Cheng Li\aff{1,2}, Yan Yang\aff{3}\corresp{\email{yanyang@udel.edu}}, Kangcheng Zhou\aff{1,2,4}, William H. Matthaeus\aff{3},
 \and Minping Wan\aff{1,2} 
 \corresp{\email{wanmp@sustech.edu.cn}}}

\affiliation{\aff{1}Guangdong Provincial Key Laboratory of Turbulence Research and Applications, Department of Mechanics and Aerospace Engineering, Southern University of Science and Technology, Shenzhen 518055, PR China
\aff{2}Guangdong-Hong Kong-Macao Joint Laboratory for Data-Driven Fluid Mechanics and Engineering Applications, Southern University of Science and Technology, Shenzhen 518055, PR China
\aff{3}Department of Physics and Astronomy, University of Delaware, DE 19716, USA
\aff{4}Department of Mechanical Engineering, The University of Hong Kong, 999077, China}

\begin{document}

\maketitle

\begin{abstract}
The Kolmogorov-Yaglom (third-order) law, %also called as third-order law, 
links energy transfer rates in the inertial range of magneto-hydrodynamic (MHD) turbulence with third-order structure functions. Anisotropy, a typical property in the solar wind, largely challenges the applicability of the third-order law with isotropic assumption. 
% A direction-averaged form of this law over a spherical surface was proposed in literature without the necessity of isotropy. 
To shed light on the energy transfer process in the presence of anisotropy, the present study conducted direct numerical simulations (DNSs) on forced MHD turbulence with normal and hyper-viscosity under various strengths of the external magnetic field ($B_0$), and calculated three forms of third-order structure function with or without averaging azimuthal or polar angles to $B_0$ direction. Correspondingly, three forms of estimated energy transfer rates were studied systematically with various $B_0$. The result shows that the peak of the estimated longitudinal transfer rate occurs at larger scales as closer to the $B_0$ direction, and its maximum shifts away from the $B_0$ direction at larger $B_0$. Compared with normal viscous cases, hyper-viscous cases can attain better separation of the inertial range from the dissipation range, thus facilitating the analyses of the inertial range properties and the estimation of the energy cascade rates. 
The direction-averaged third-order structure function over a spherical surface proposed in literature predicts the energy transfer rates and inertial range accurately, even at very high $B_0$. With limited statistics, the calculation of the third-order structure function shows a stronger dependence on averaging of azimuthal angles than the time, especially at high $B_0$ cases.  These findings provide insights into the anisotropic effect on the estimation of energy transfer rates.
\end{abstract}
% , thus technically requiring more time frames and more complete coverage of angles for convergence
%\begin{keywords}
%Authors should not enter keywords on the manuscript, as these must be chosen by the author during the online submission process and will then be added during the typesetting process (see http://journals.cambridge.org/data/\linebreak[3]relatedlink/jfm-\linebreak[3]keywords.pdf for the full list)
%\end{keywords}

\section{Introduction}
Magneto-hydrodynamic (MHD) turbulence commonly exists in nature, such as the solar wind with high Reynolds numbers  \citep{Coleman68,BrunoCarboneLRSP13,JokipiiHollweg70,MattGold82a,TuMarsch95,Parker-cmf}, on which we focus here. From the engineering perspective, the solar wind has an influence on the weather in space, where it impacts the functioning of satellites. For fundamental research, the cross-scale energy transfer is an important process for the analysis and modelling of MHD turbulence. In the classical energy cascade scenario, energy is transferred from large to small scales at a constant rate, which is finally dissipated at the dissipation range \citep{Taylor38,KarmanHowarth38,Kol41a,Kol41b,Kraichnan71-jfm}. This picture has been adapted to MHD turbulence \citep{hossain1995phenomenology,politano1998karman}. 
It is trivial to obtain the dissipation rate in terms of \textit{ad hoc} viscosity and resistivity, as implemented for example, in MHD simulations. However, space plasmas often behave as collisionless plasmas, for which the classical viscosity and resistivity become inapplicable, and therefore also inapplicable is the viscous and resistive dissipation rate.
In the absence of a simple closed expression for dissipation function, there has been increasing interest in resorting to the cross-scale energy transfer process to quantify the energy transfer rate in the inertial range (sometimes loosely called cascade rate). For example, starting from the von Kármán-Howarth (vKH) equation \citep{de1938statistical,monin1975statistical,FrischBook95}, a four-fifths law, originally derived by Kolmogorov in hydrodynamic turbulence assuming isotropy, links the dissipation rate with the third-order moment of longitudinal velocity increments \citep{kolmogorov1941dissipation}, i.e. $\langle  |\delta u_l|^3 \rangle$. This law was modified in a slightly more general form, a four-thirds law \citep{monin1975statistical,frish1995turbulence,antonia1997analogy}, by replacing the second-order moment of the longitudinal velocity increments with the sum of the square of the three velocity increments, i.e. $\langle \delta u_l |\delta\boldsymbol{u}|^2 \rangle$. An analogy of this law, also called the third-order law here, in incompressible MHD turbulence was derived by \cite{politano1998karman} under the assumption of isotropy.

Given that turbulence is frequently simplified, e.g., assumed to be isotropic and incompressible, in most treatises on the energy transfer process, it is natural to enquire about the effects introduced by implementing other realistic complexities. For example, the isotropic third-order law has been generalized to take into account corrections from anisotropy  \citep{Osman11-ApJ,Stawarz2011third,Podesta2008laws,verdini2015anisotropy}, compressibility \citep{YangEA-POF-17,Hadid2017energy,Kritsuk09,Carbone09,Banerjee16b,AndresEA19}, solar wind shear \citep{Wan2009third,Wan2010third} and expansion \citep{Gogoberidze2013yaglom,Hellinger2013proton}, and Hall effect \citep{Hellinger2018karman,Ferrand2019exact,ferrand2022depth,BandyopadhyayEA20-Hall}.
Here we seek to systematically investigate the effect of anisotropy on the cross-scale energy transfer in the inertial range. Anisotropy is inherent in turbulence threaded by a guiding magnetic field $\boldsymbol{B}_0$ (e.g. \citep{shebalin1983anisotropy,MattEA96-var,HorburyEA08,OughtonEA15_anisotropy}), as in the solar wind. It is widely accepted that the cross-scale energy transfer is suppressed along the parallel direction with respect to the mean magnetic field, which has been shown explicitly in terms of third-order structure functions using MHD turbulence simulations \citep{verdini2015anisotropy}. By computing the divergence of third-order structure functions along different directions of lags, \cite{verdini2015anisotropy} characterizes the anisotropy of energy transfer and the so-obtained transfer rate depends on the angle between $\boldsymbol{B}_0$ and the direction of lags. Therefore, the isotropic third-order law, although widely used in solar wind studies, is seriously flawed in that it does not take into account the angular dependence of energy transfer in anisotropic MHD turbulence. 
The presence of this 
anisotropy impacts in particular
experimental estimations of energy transfer, 
as exemplified by the upcoming Helioswarm solar wind
mission \citep{SpenceEA19,MatthaeusEA19-whitepaper}.  %\includegraphics{Figures_jfm_22_11_2022_JPG/Current_hyper2e7_norm8e4Contour_19_10.jpg} % Bin: a typo by Bill? 

To expand the applicability of the third-order law for anisotropic MHD turbulence, the most straightforward way to proceed would be to directly compute the divergence of the energy-flux vector. The energy-flux vector is actually the third-order structure function, and its projection along the lag direction is used in the isotropic third-order law, wherein the energy-flux vector in the inertial range is nearly radial in lag space. However, an accurate determination of this divergence form requires information at all points in 3D lag space, necessitating simultaneous multi-point measurements that span 3D spatial directions. This is obviously not feasible with single-spacecraft data and even with multi-spacecraft data due to the small number of available lag directions. To overcome the difficulty, \cite{podesta2007anisotropic} and \cite{galtier2009exact} modified the isotropic third-order law with external $\boldsymbol{B}_0$, employing additional assumptions. 
But we do not implement these theories due to their general complexity. The divergence form of the energy-flux vector can be simplified under certain symmetry. For example, in the rotating turbulence having azimuthal symmetry with respect to the rotational axis \citep{yokoyama2021energy} and the anisotropic MHD turbulence having azimuthal symmetry with respect to the guiding magnetic field \citep{Alexakis07a}, the divergence form  can be simplified by integration over the azimuthal angle. Another simplification was realized originally in hydrodynamic turbulence by solid angle averaging over all possible orientations of the lag vector \citep{NieTanveer99,TaylorEA03}, which was then adapted in MHD turbulence \citep{Wan2009third,Osman11-PRL}. Recently, \cite{wang2022strategies} investigated such a directional average of the third-order law over a number of lag directions on a spherical surface. In comparison with the isotropic third-order law, this direction-averaged version attains more accurate energy dissipation rate and will be called direction-averaged third-order law hereafter. 

These preliminary demonstrations provide supporting but incomplete evidence to develop a discrete formulation that is representative of the anisotropic energy transfer process and is applicable in both numerical analysis and in observational realizations
such as Helioswarm
\citep{SpenceEA19}. 
To advance these issues, here we conduct a systematic study of the angular dependence of the third-order law and the effect of the number of samples over directions spanning solid angle with various strengths of external mean magnetic field ($B_0$). Besides the direction averaging, the effect of time averaging is also investigated. 
Concerning the energy cascade process in the inertial range, one requires the existence of a range of scales, in which the dynamics is dominated by inertia terms and is well separated from both 
the energy-containing range and the dissipation range. Hyper-viscosity is often used to extend the inertial range, which attains a similar Reynolds number with lower computational costs. \cite{spyksma2012quantifying} compared the characteristics of the normal with hyper-viscous simulations, and formulated the characteristic lengthscale and Reynolds number for the hyperviscous case. 
\cite{biskamp2000scaling} conducted isotropic MHD simulation with hyperviscosity to attain an elongated  inertial range well separated from the dissipation range. They reported that the bottleneck effect is invisible in structure function profiles, but can be identified in energy spectra, introducing a hump at the end of the inertial range. \cite{beresnyak2009comparison} simulated anisotropic MHD turbulence with hyperviscosity and various $B_0$, and claimed that the bottleneck effect is inhibited by external magnetic fields in energy spectra. Here we also conduct numerical simulations of anisotropic MHD turbulence with hyperviscosity, but the emphasis is on the impact of hyperviscosity on the evaluation of the third-order law with varying external magnetic fields. 

The structure of the paper is as follows: In Section \ref{sec:numerical_methods}, the numerical method will be introduced, including the governing equations, simulation configurations and several characteristics. A brief review of the third-order law is given in Section \ref{sec:theory}. In Section \ref{sec:results}, the effects of hyperviscosity on structure functions are discussed, and the effects of directional and time averaging on the third-order law will be given. The key findings will be listed in the conclusion. %In the appendix, the simulation convergence is illustrated.
 
\section{Numerical methods}\label{sec:numerical_methods}
\subsection{Governing Equations}

The hyperviscosity modified governing equation for the simulation of incompressible MHD turbulence is written as \cite{biskamp2003magnetohydrodynamic}:
\begin{equation}
\frac{\partial \boldsymbol{v}} {\partial t} + (\boldsymbol{v} \cdot \nabla)\boldsymbol{v}  = -\nabla (p + \frac{|\boldsymbol{b}|^2}{2}) + (\boldsymbol{b} \cdot \nabla)\boldsymbol{b} + (\boldsymbol{B_0} \cdot \nabla)\boldsymbol{b} + (-1)^{h+1} \nu_h \nabla^{2h}\boldsymbol{v} + \boldsymbol{f_v}\ , 
\label{eq:governeq_velocity}
\end{equation}
\begin{equation}
\frac{\partial \boldsymbol{b}} {\partial t} + (\boldsymbol{v} \cdot \nabla)\boldsymbol{b}  =  (\boldsymbol{b} \cdot \nabla)\boldsymbol{v} + (\boldsymbol{B_0} \cdot \nabla)\boldsymbol{v} + (-1)^{h+1} \eta_h \nabla^{2h}\boldsymbol{b} \ ,
\label{eq:governeq_magnetic}
\end{equation}
\begin{equation}
 \nabla \cdot \boldsymbol{v} =0\ ,
\label{eq:icmpV}
\end{equation}
\begin{equation}
\nabla \cdot \boldsymbol{b} =0 \ ,
\label{eq:singB}
\end{equation}
where $\boldsymbol{v}$ and $\boldsymbol{b}$ represent the velocity vector and the fluctuating magnetic field, respectively. An external mean magnetic field, $\boldsymbol{B}_0 = B_0 \boldsymbol{e_z}$, is imposed along $z$-direction. Its magnitude is normalized by the Root-Mean-Square (RMS) of magnetic fluctuation, $b_{rms} (\approx 1)$. $p$ is the thermal pressure;
$\nu_h$ and $\eta_h$ denote the kinetic viscosity and the magnetic resistivity coefficients, respectively. The power $h$ is the order of hyperviscosity, where $h=1$ represents normal viscosity and $h>1$ represents hyperviscosity. An external force, $\boldsymbol{f_v}$, is added to the kinetic governing equation to achieve a statistically stationary state. The forcing is solenoidal to avoid introducing compression into the velocity field.

\subsection{Configuration setup}
We solve the Fourier-space version of the governing equations via the pseudo-spectral method with de-aliasing by the two-thirds rule. The computational domain is a cube of dimension $[0,2\pi]^3$ with periodic boundary conditions in all directions. The 2\ts{nd}-order Adam-Bashforth scheme is employed for time advancement. %All runs discussed here are driven problems. mpw: this is repeating and unnecessary wording.
The external force, $\boldsymbol{f_v}$, acts only on the first two wavenumber shells, i.e. $k=1$ and $k=2$, without affecting the inertial range, where $k$ is the norm of the wavenumber vector, $\boldsymbol{k}$. The forcing is achieved by keeping the constant energy injection rate. 
Mathematically, it is a random, Gaussian-distributed and $\delta(t)$-correlated function as in \cite{yang2021effects}. All runs are initialized with random velocity and magnetic fluctuations within the wavenumber band $k\in[1,~5]$, with spectra proportional to $1/[1+(k/k_0)]^{11/3}$ and $k_0=3$. The initial kinetic and magnetic energies are equal, i.e. $E_v=E_b=0.5$. The cross helicity is almost zero.  Equal viscosity and resistivity (i.e., the magnetic Prandtl number is unity) are used for all simulations.

To study the anisotropic energy transfer process in the inertial range, we initialize our simulations with a range of external mean magnetic fields, $B_0$, and two types of viscosity. These runs are grouped into two series and more details are listed in Table \ref{table:setup}. The first series of runs are conducted in 1024$^3$ grids using normal viscosity (i.e., $h=1$), $\nu_h=8 \times 10^{-4}$, and the second series of runs are conducted in $512^3$ grids using hyperviscosity at the order of $h=2$, $\nu_h=2 \times 10^{-7}$. All simulations are fully resolved with $k_{max} \eta_{k,v}>1.5$, where $k_{max}$ is the largest resolved wavenumber.
The time to reach a statistically stationary state and the sampling period are listed in Table \ref{table:setup}, and all cases are sampled with 0.5 large-eddy turnover time ($T_e$). Throughout the paper, the time will be in units of the large-eddy turnover time $T_e$ at $B_0=0$. %{\color{red}normal or hyper-viscous? they are almost identical with 1.01 and 1.04 for normal and hyper-viscous cases, respectively.}. 
The cases without or with a relatively weak external mean magnetic field, i.e., $B_0=$0 and 2, reach the statistically stationary state earlier than the higher $B_0$ cases. To obtain the following statistical properties, we use 20 time frames over 10 $T_e$ for the normal viscous cases, and for the hyper-viscous cases, 30, 120 and 150 time frames over 15$T_e$, 60$T_e$ and 75$T_e$ at $B_0=$ 0, 2 and 5, respectively.% Due to the less grid points , the averaging period for the hyper-viscous cases increases to 15$T_e$, 60$T_e$ and 75$T_e$ with 30, 120 and 150 time frames at $B_0=$ 0, 2 and 5, respectively.}
% The computational cost for the normal-viscous case is about 5 times }% For these smaller $B_0$ cases, the time to reach a statistically stationary state is about 5$T_e$. }
%The convergence of these simulations is shown in the Appendix A by plotting the temporal evolution of the total energy and the dissipation rate. The specific convergent period,} starting with the stable evolution of the total energy and ending up with the final instant of the simulation, 

More observables for all simulations are listed in Table \ref{table:setup} and they are time-averaged. The kinetic dissipation rate, $\varepsilon_{v}$, and the magnetic dissipation rate, $\varepsilon_{b}$, are calculated as:
\begin{equation}
    \varepsilon_{v} = \nu_h \sum\limits_{\boldsymbol{k}} k^{2h} \langle |\boldsymbol{\hat{u}} (\boldsymbol{k},t)|^2 \rangle \ , \; \;
    \varepsilon_{b} = \eta_h \sum\limits_{\boldsymbol{k}} k^{2h} \langle |\boldsymbol{\hat{b}} (\boldsymbol{k},t)|^2 \rangle  \ ,
\label{eq:dissB}
\end{equation}
where $\boldsymbol{\hat{u}}$ and $\boldsymbol{\hat{b}}$ denote the velocity and magnetic vectors in the Fourier space. The operator $\langle \cdot \rangle$ denotes the ensemble average which is identical to the space average for homogeneous turbulence. The kinetic dissipation rate increases with $B_0$, while the magnetic dissipation rate decreases. 
To quantify the anisotropy at different $B_0$, the variable, $\theta_{v}$, originally introduced by  \cite{shebalin1983anisotropy}, is used, $\tan^2 \theta_v = {\sum k_\perp^2 |\boldsymbol{u}(\boldsymbol{k},t)|^2}/{\sum k_z^2 |\boldsymbol{u}(\boldsymbol{k},t)|^2}$.
In the isotropic case, $\theta_{v}$ equals 54$^\circ$. For an extreme case with all energy in the perpendicular plane to $B_0$ direction, $\theta_{v}$ is close to 90$^\circ$. Table 1 shows that $\theta_{v}$ is larger  at a higher $B_0$ value.

\begin{table}
  \begin{center}
   \def~{\hphantom{0}}
	\begin{tabular}{lccccccccc}
		$B_0$ & $h$ & Grids  &  $\varepsilon_v$ & $\varepsilon_b$ & $R_{\lambda,v}$ & $R_{\lambda,b}$ & $k_{max} \eta_{k,v}$ & $\theta_{v}(^\circ)$ & Averaging period ($T_e$)        \\ [3pt]
        0 & 1 &	1024$^3$ & 0.59 & 1.39 & 290 & 89 & 1.86 & 62& [5:19] \\  
        2 & 1 &	1024$^3$ & 0.78 & 1.07 & 295 & 164 & 1.73 & 73& [5:17]\\  
        5 & 1 &	1024$^3$ & 0.81 & 0.80 & 467 & 166 & 1.71 & 83& [9:18]\\ 
        0 & 2 &	512$^3$ & 0.67 & 1.24 & 846 & 255 & 1.73 & 55& [15:30]\\   
        2 & 2 &	512$^3$ & 0.77 & 1.12 & 951 & 473 & 1.71 & 72& [8:68]\\ 
        5 & 2 &	512$^3$ & 0.83 & 1.02 & 2415 & 435 & 1.70 & 83& [200:275]\\  
        %15 & 2 & 512^3 & 0.98 & 0.80 & 15681 & 534 & 1.67 & 87& [60:122]\\ \hline
	\end{tabular}
	\caption{Configuration setup of normal-viscous and hyper-viscous simulations at different external mean magnetic fields ($B_0$). $h$ denotes the order of the hyperviscosity; $\varepsilon_{v}$ and $\varepsilon_{b}$ represent the kinetic and magnetic dissipation rates, respectively; $R_{\lambda,v}$ and $R_{\lambda,b}$ are the kinetic and magnetic Taylor Reynolds numbers, respectively; $k_{max} \eta_{k,v}$ illustrates the grid resolution, where $\eta_{k,v}$ and $k_{max}$ are respectively the Kolmogorov lengthscale and the maximum resolved wavenumber (a third of the total grid in one direction); $\theta_v$ describes the anisotropic intensity as in \cite{shebalin1983anisotropy}; $T_e$ refers to the large-eddy turnover time at $B_0=$ 0.}
	\label{table:setup}
  \end{center}
\end{table}

\begin{figure}
	\centering
	\includegraphics[width=0.9\linewidth,keepaspectratio]{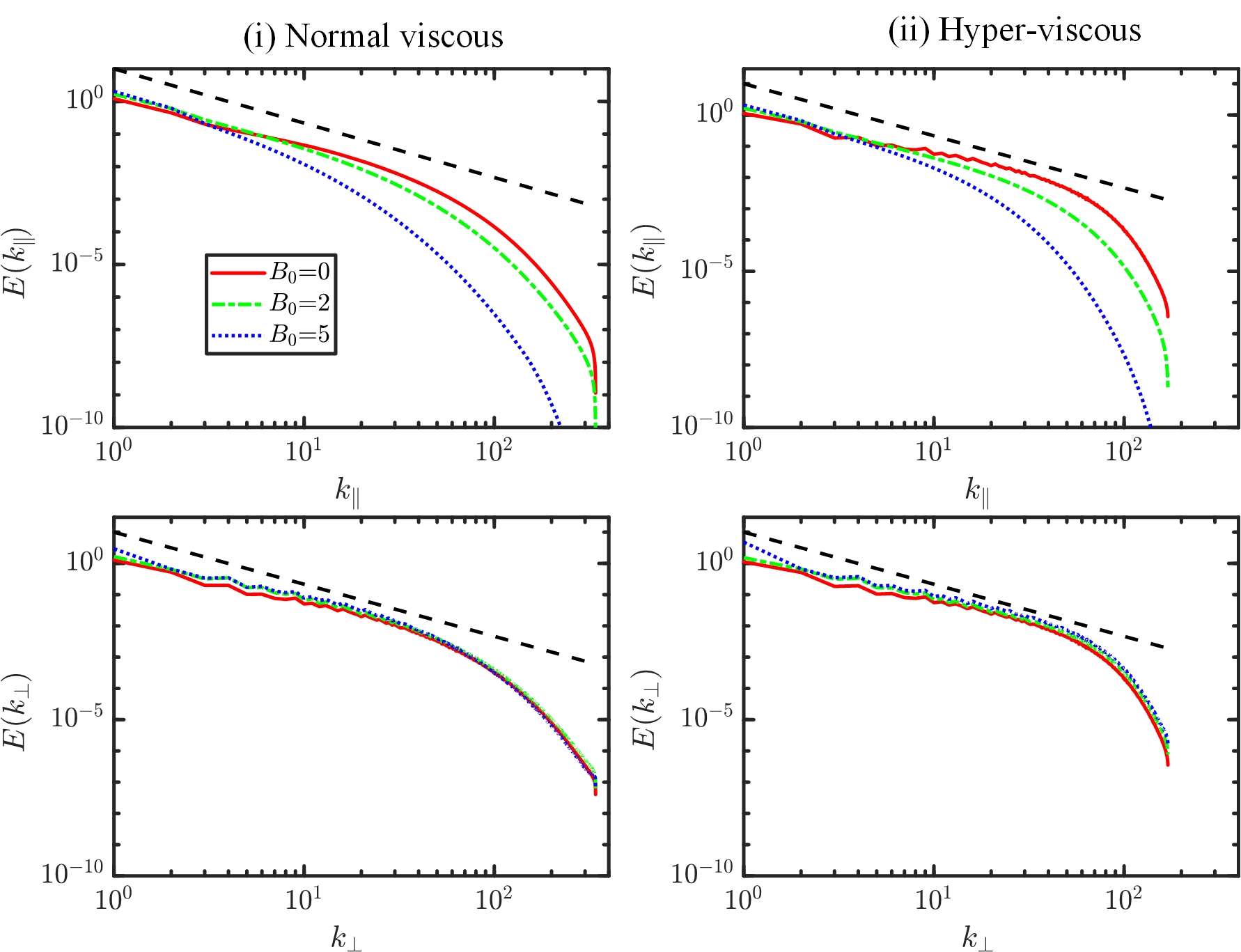}	\\[0.2cm]
	\caption{Parallel (top) and perpendicular (bottom) reduced spectra of total energy $E=E_v+E_b$, $E(k_{\parallel})$ and $E(k_{\perp})$ in normal viscous simulations (left) and hyper-viscous simulations (right). Dashed black lines show $k^{-5/3}$ power law for reference.} %{\color{red}modify this figure}}
	\label{fig:ParPerES_B5rnu2e7_V1}
\end{figure}
Figure \ref{fig:ParPerES_B5rnu2e7_V1} shows the reduced spectra of total energy $E=E_v+E_b$ in the $k_\parallel$ and $k_\perp$ plane for the normal and hyper-viscous simulations, where $k^{-5/3}$ power law is also shown for reference. The parallel and perpendicular spectra are defined as $E(k_\parallel)=\int E(k_\perp, k_\parallel)dk_\perp$ and $E(k_\perp)=\int E(k_\perp, k_\parallel)dk_\parallel$, respectively. It is clear that the parallel spectrum is suppressed and the perpendicular spectral transfer is stronger than the parallel transfer with increasing $B_0$.
This reflects the decreasing of the dissipation rate with the increasing of $B_0$ as listed in Table \ref{table:setup}. %JB: can it indicate the change of the dissipation rate?
One may observe that higher energy resides in the first $\sim$3 wave modes at larger $B_0$ as reported in Ref. \citep{alexakis2007anisotropic}. The inertial range is roughly viewed as the range of scales over which the spectrum fits well with the $k^{-5/3}$ power law. As expected, the simulations with hyperviscosity realize an elongated inertial range.

\section{Direction-averaged third-order law}\label{sec:theory}
This section describes three types of averaging relevant to the third-order law. Starting from the von Kármán-Howarth (vKH) equation \citep{de1938statistical,monin1975statistical,frish1995turbulence}, a general form of the third-order law will be derived. Considering that the present study relates to the
effects of an externally supported  mean magnetic field, the three types of averaging applied to the general form will be discussed under different assumptions regarding 
anisotropy.

The vKH equation is typically composed of the time-rate-of-change of energy, energy transfer across scales, and energy dissipation terms, which respectively dominate at energy injection, inertial and dissipation scales.
These contributions are evident in the
vKH equation itself,  % do not understand, does it mean from the math expression directly?
\begin{equation}
\frac{\partial}{\partial t} \langle (\delta\boldsymbol{z}^{\pm})^2 \rangle =
- \nabla_l \cdot \langle \delta\boldsymbol{z}^{\mp} |\delta\boldsymbol{z}^{\pm}|^2 \rangle + 
2\nu \nabla^2_l \langle (\delta\boldsymbol{z}^{\pm})^2 \rangle  -4 \varepsilon^{\pm}.  
\label{eq:VkH_eq}
\end{equation}
Here $\nabla_l$ denotes derivatives with respect to the lag vector $\boldsymbol{l}$; $\boldsymbol{Y}^{\pm}=\langle \delta\boldsymbol{z}^{\mp} |\delta\boldsymbol{z}^{\pm}|^2 \rangle $ is the third-order structure function, also called as (Yaglom) energy-flux vector with the Els\"asser variable ($\boldsymbol{z}^{\pm} = \boldsymbol{u} \pm \boldsymbol{b}$) increment being defined as $\delta\boldsymbol{z}^{\pm} (\boldsymbol{x},\boldsymbol{l})= \boldsymbol{z}^{\pm} (\boldsymbol{x}+\boldsymbol{l}) - \boldsymbol{z}^{\pm} (\boldsymbol{x})$.
$\varepsilon^{\pm}=\nu_h  \sum\limits_{\boldsymbol{k}} k^{2h} \langle |\hat{\boldsymbol{z}}^{\pm} (\boldsymbol{k},t)|^2 \rangle $ represent the mean dissipation rates of Els\"asser energies.

In the inertial range with negligible contribution from non-stationary and dissipative
terms in Eq. (\ref{eq:VkH_eq}), the cross-scale energy transfer is expressed as
\begin{equation}
\nabla_l \cdot \boldsymbol{Y}^{\pm} = \nabla_l \cdot \langle \delta\boldsymbol{z}^{\mp} |\delta\boldsymbol{z}^{\pm}|^2 \rangle = -4 \varepsilon^{\pm}, 
\label{eq:div_3rd_order_law}
\end{equation}
Note that after ensemble averaging, no dependence of $\boldsymbol{Y}^{\pm}$ on the position $\boldsymbol{x}$ remains, and $\varepsilon^{\pm}$ is independent of both position $\boldsymbol{x}$ and lag $\boldsymbol{l}$.

% Given the isotropic condition, the Eq.(\ref{eq:div_3rd_order_law}) can be written as:
% \begin{equation} 
% \nabla_l \cdot (Y_l^{\pm} \frac{\boldsymbol{l}}{l})  = Y_{l,isotropic}^{\pm} \nabla_l \cdot \frac{\boldsymbol{l}}{l} = Y_{l,isotropic}^{\pm} \frac{3}{l} = -4 \varepsilon^{\pm} 
% \label{eq:iso_3d_3rd_order_law01}
% \end{equation} 
% where $Y_l^{\pm} =\langle \delta z_l^{\mp} |\delta\boldsymbol{z}^{\pm}|^2 \rangle$ is the projection of the energy flux vector along $\boldsymbol{l}$; $\delta z_l^{\mp} = \delta\boldsymbol{z}^{\mp} \cdot \frac{\boldsymbol{l}}{l}$, and $l$ is the norm of $\boldsymbol{l}$. Equation (\ref{eq:iso_3d_3rd_order_law01}) can then be recast into the classical isotropic third-order law as follows:
% \begin{equation} 
% Y_{l,isotropic}^{\pm}  = -\frac{4}{3} \varepsilon^{\pm} l, 
% \label{eq:iso_3d_3rd_order_law02}
% \end{equation} 

In anisotropic conditions, as in the present study with imposed external mean magnetic field, Eq. (\ref{eq:div_3rd_order_law}) can be reformed by taking a volume integral in a sphere of radius $l$ as follows:
\begin{equation}
\iiint_{|\boldsymbol{l}|\le l} \nabla_l \cdot \boldsymbol{Y}^{\pm} dV= \iiint_{|\boldsymbol{l}|\le l} -4 \varepsilon^{\pm} dV. 
\label{eq:div02_3rd_order_law}
\end{equation}
Using Gauss's theorem, Eq.(\ref{eq:div02_3rd_order_law}) can be written as a surface integral
\begin{equation}
\oint_{|\boldsymbol{l}|=l} Y_l^{\pm} dS=  \iiint_{|\boldsymbol{l}|\le l} -4 \varepsilon^{\pm} dV = -\frac{16 \pi}{3} \varepsilon^{\pm} l^3, 
\label{eq:div03_3rd_order_law}
\end{equation}
where $Y_l^{\pm} =\langle \delta z_l^{\mp} |\delta\boldsymbol{z}^{\pm}|^2 \rangle$ is the projection of the energy-flux vector along $\boldsymbol{l}$; $\delta z_l^{\mp} = \delta\boldsymbol{z}^{\mp} \cdot \frac{\boldsymbol{l}}{l}$, and $l$ is the norm of $\boldsymbol{l}$.
In spherical coordinates, Eq.(\ref{eq:div03_3rd_order_law}) can be written as:
\begin{equation}
\frac{1}{4\pi}\int_{0}^{2\pi} \int_{0}^{\pi} Y_l^{\pm} \sin\theta d\theta d\phi = -\frac{4}{3} \varepsilon^{\pm} l, 
\label{eq:3d_3rd_order_law}
\end{equation}
where $\theta$ represents the polar angle and $\phi$ the azimuthal angle. Note that this integral form Eq.~(\ref{eq:3d_3rd_order_law}) is identical to the derivative form Eq.~(\ref{eq:div_3rd_order_law}), but it is simpler in the sense that accurate determination of the integration only requires information on the 
spherical surface spanned by the $(\theta, \phi)$ coordinates in the 3D lag space. (See e.g.,  \cite{TaylorEA03,wang2022strategies})

The most general form of $Y_l^{\pm}$ should be a function of $l$, $\theta$, and $\phi$, i.e., $Y_l^{\pm}(l,\theta,\phi)$. However, there is no universal expression of $Y_l^{\pm}(\theta,\phi)$ so far, as little information is available on its variation with turbulence parameters. Previous studies have either been limited to the purely isotropic assumption (i.e., $Y_l^{\pm}$ is independent of $\theta$ and $\phi$) or treated anisotropic turbulence with azimuthal symmetry(i.e., $Y_l^{\pm}$ is independent of $\phi$) as implemented, for example
by \cite{Stawarz09}. %The former one can be attained from 
Under isotropic assumption, Eq.(\ref{eq:3d_3rd_order_law}) can be reduced to:
\begin{equation}
\frac{1}{4\pi} \; Y_{l,{\rm isotropic}}^{\pm}\int_{0}^{2\pi} \int_{0}^{\pi}  \sin\theta d\theta d\phi = Y_{l,{\rm isotropic}}^{\pm} = -\frac{4}{3} \varepsilon^{\pm} l. 
\label{eq:iso_3rd_order_law}
\end{equation}

To better understand the anisotropic energy transfer in the inertial range, here we provide a systematic study of $Y_l^{\pm}$'s dependence on $\theta$ and $\phi$ with different guide field magnitudes. Specifically, three forms of $Y_l^{\pm}$ are discussed:

I) The general form of the third-order structure function for every lag vector $\boldsymbol{l}=(l, \theta, \phi)$ in 3D lag space,
\begin{equation}
 Y_l^{\pm}(\theta,\phi)=\langle \delta z_l^{\mp} |\delta\boldsymbol{z}^{\pm}|^2 \rangle=Y_l^{\pm}(\theta_i,\phi_j),
\label{eq:3rd_order_sf_3D_lag}
\end{equation} 
% describes every detail of the radial or longitudinal energy transfer, varying in both azimuthal $\phi$ and polar $\theta$ directions. 
represents a local radial or longitudinal energy transfer, and `local' means at the specific azimuthal and polar angles, while the total radial energy transfer is the sum of the contributions, i.e. Eq.(\ref{eq:3rd_order_sf_3D_lag}), from all azimuthal $\phi$ and polar $\theta$ directions at the same lag. 
Due to the axisymmetric external mean magnetic field, the range of $\theta$ is $[0^\circ, 90^\circ]$. Lag vectors in 37 directions (there is only one direction at $\theta=0^\circ$), uniformly spaced in azimuthal and polar angles ($\Delta\theta=15^\circ$ and $\Delta\phi=60^\circ$), are used to cover the sphere. A 3D Lagrangian interpolation was used to attain the data not located at grid points. Separate estimates are made for each of 37 directions, i.e., $Y_l^{\pm}(\theta_i,\phi_j)$, $\theta_i\in[0^\circ:15^\circ:90^\circ]$ and $\phi_j\in[0^\circ:60^\circ:300^\circ]$.

II) The azimuthal averaged form of the third-order structure function,
\begin{equation}
\widetilde{ Y_l^{\pm}}(\theta) = \frac{1}{2\pi}\int_{0}^{2\pi} Y_l^{\pm}(\theta,\phi) d\phi
\approx \frac{\sum_{j=1}^{N_j} Y_l^{\pm}(\theta_i,\phi_j) }{N_j},
\label{eq:3rd_order_sf_2D_lag}
\end{equation}
% describes the anisotropy of energy transfer in the radial direction at a fixed polar ($\theta$) direction. 
describes the anisotropy of local radial energy transfer, and `local' means at a specific polar angle, while the total transfer rate is the sum of the contributions, i.e. Eq.( \ref{eq:3rd_order_sf_2D_lag}), from all polar $\theta$ directions at the same lag.
$N_j$(=6) represents the number of azimuthal angles. Following Eq.~(\ref{eq:3rd_order_sf_3D_lag}), separate estimates are made for each of 37 directions, and then averaged over 6 azimuthal directions.

III) The direction-averaged form of the third-order structure function,
\begin{equation}
\overline{ Y_l^{\pm}}=\frac{1}{4\pi}\int_{0}^{2\pi} \int_{0}^{\pi} Y_l^{\pm} \sin\theta d\theta d\phi \approx \frac{\sum_{j=1}^{N_j} \sum_{i=1}^{N_i}  Y_l^{\pm}(\theta_i,\phi_j) \sin\theta_i}{N_j\sum_{i=1}^{N_i} \sin\theta_i},
\label{eq:3rd_order_sf_1D_lag}
\end{equation}
is rather general since it takes into account all possible anisotropy in both azimuthal and polar directions. $N_i$(=7) and $N_j$(=6) represent the number of angle $\theta$ and $\phi$. %As indicated in Eq.(\ref{eq:3d_3rd_order_law}), this direction-averaged form only depends on the lag length $l$, which presumably provides an accurate estimation of energy cascades rate, $\varepsilon_{diss}$.As a guidance for the physical interpretation of the above three forms in the following analysis on the cases with external mean magnetic fields,
This direction-averaged form of the third-order structure function is derived directly from the vKH equation without the assumption of isotropy and only depends on the lag length $l$. After normalizing $\overline{ Y_l^{\pm}}$ with the lag $-4/3l$, it can represent an accurate estimation of the cross-scale energy transfer rate (or energy dissipation rate $\varepsilon_{diss}$) and the inertial range. %and  energy cascades rate, $\varepsilon_{diss}$.}

% {\color{blue} the other two forms only represent an estimated radial or transfer rate, and `local' in the sense of a fixed polar angle or both , while the total transfer rate is the sum of the contributions from all polar angles at the same lag. Specifically, the $\widetilde{ Y_l^{\pm}}/(-4/3l)$ represents the estimated `local' transfer rate. Similarly, the $ Y_l^{\pm}/(-4/3l)$ represents the estimated local transfer rate at a fixed both azimuthal and polar angles. }

% {\color{blue} As a guidance for the physical interpretation of the above three forms in the following analysis on the cases with external mean magnetic fields, the direction-averaged form, $\overline{ Y_l^{\pm}}$ in Eq.(\ref{eq:3rd_order_sf_1D_lag}), is derived directly from the vKH equation without the assumption of isotropy and only depends on the lag length $l$. After normalizing $\overline{ Y_l^{\pm}}$ with the lag $-4/3l$, it can represent an accurate estimation of cross-scale energy transfer rate, also an estimation for the inertial range and  energy cascades rate, $\varepsilon_{diss}$. 
% However, the other two forms only represent an estimated radial or transfer rate, and `local' in the sense of a fixed polar angle or both , while the total transfer rate is the sum of the contributions from all polar angles at the same lag. Specifically, the $\widetilde{ Y_l^{\pm}}/(-4/3l)$ represents the estimated `local' transfer rate. Similarly, the $ Y_l^{\pm}/(-4/3l)$ represents the estimated local transfer rate at a fixed both azimuthal and polar angles. }

The aforementioned description of the energy transfer in the three forms of third-order structure functions can provide estimations of the actual energy transfer rate to different degrees. %they are also the different approaches to estimate the energy cascade rates. 
For instance,  Eq.(\ref{eq:3rd_order_sf_3D_lag}) has been widely used in the observational measurements with one spacecraft. To clearly show the estimation of the energy transfer rate and the inertial range, the aforementioned three forms of the third-order structure functions, i.e., $ Y_l^{\pm}$, $\widetilde{ Y_l^{\pm}}$  and $\overline{ Y_l^{\pm}}$, will be averaged on their $+$ and $-$ components and normalized with the lag and the actual cascade rate, i.e. $-\frac{4}{3} \varepsilon_{diss} l $, where $\varepsilon_{diss}$ is the total cascade rate, $\varepsilon_{diss}= (\varepsilon^+ + \varepsilon^-)/2 =  \varepsilon_b + \varepsilon_v$.%i) Without assuming any symmetry, Eq.~\ref{eq:3d_3rd_order_law} is rather general since it takes into account all possible anisotropy in both azimuthal and polar directions. Providing that the intervals of the polar and azimuthal angles are equal, a discrete form of Eq.(\ref{eq:3d_3rd_order_law}) can be written as
%\begin{equation}
%\overline{ Y_l^{\pm}} = \frac{\sum_{j=1}^{N_j} \sum_{i=1}^{N_i}  Y_l^{\pm}(\theta_i,\phi_j) sin\theta_i}{N_j\sum_{i=1}^{N_i} sin\theta_i } = -\frac{4}{3} \varepsilon^{\pm} l,
%\label{eq:3Ddiscrete_3rd_order_law}
%\end{equation}
%where $N_i$ and $N_j$ represent the number of angle $\theta$ and $\phi$, respectively. Due to the axisymmetric external mean magnetic field, the range of $\theta$ is $[0^\circ, 90^\circ]$. A 3D Lagrangian interpolation was used to attain the data not located at grid points. $\overline{ Y_l^{\pm}}$ depends on the lag length $l$. 

%ii) Assuming the isotropic in the perpendicular plane to the $B_0$, after doing an average on Eq.(\ref{eq:3Ddiscrete_3rd_order_law}) in the azimuthal direction not in the polar direction, then it can be written as:
%\begin{equation}
%\widetilde{ Y_l^{\pm}}(\theta) = \frac{\sum_{j=1}^{N_j} Y_l^{\pm}(\theta_i,\phi_j) }{N_j} ,  \ 
%\label{eq:2Ddiscrete_3rd_order_law}
%\end{equation}

%iii) Assuming the isotropic in both the parallel and perpendicular planes to the $B_0$ and by projecting $\boldsymbol{Y}^{\pm}$ along the longitudinal direction, the 1D isotropic third-order law is attained as follows:% shown in Eq.(\ref{eq:1d_3rd_order_law})as a function of $\theta$ and $\phi$ in the polar coordinates
%\begin{equation}
% Y_{l}^{\pm} = -\frac{4}{3} \varepsilon^{\pm} l \ ,
%\label{eq:1d_3rd_order_law}
%\end{equation}

%% Results and discussion
\section{Results and Discussion}\label{sec:results}
% \FloatBarrier
This section will make a comparison between normal viscous and hyper-viscous cases and focus on the angular dependence of third-order structure functions under various strengths of the external magnetic field. In addition, the possible effect of time averaging will be discussed.
%YYang: 1. the time or time interval in each part should be indicated.
% 2. a time evolution of Ev+Eb, showing the statistically steady state would help.

\subsection{Effects of hyperviscosity and polar angle dependence}	 
\label{sec:Structure_functions}
In anisotropic MHD turbulence with a mean magnetic field, the energy transfer is often deemed to be isotropic in the plane perpendicular to $B_0$ direction, that is, $Y_l^{\pm}$ is $\phi$ independent. So as a first analysis, we integrate $Y_l^{\pm}$ over $\phi$ and time average over long statistically stationary periods and attain a normalized third-order structure function as: $-3(\widetilde{ Y_l^+}+\widetilde{ Y_l^-})/(4\varepsilon_{diss} l) $. 
 \begin{figure} 
	\centering
	\includegraphics[width=0.9\linewidth,keepaspectratio]{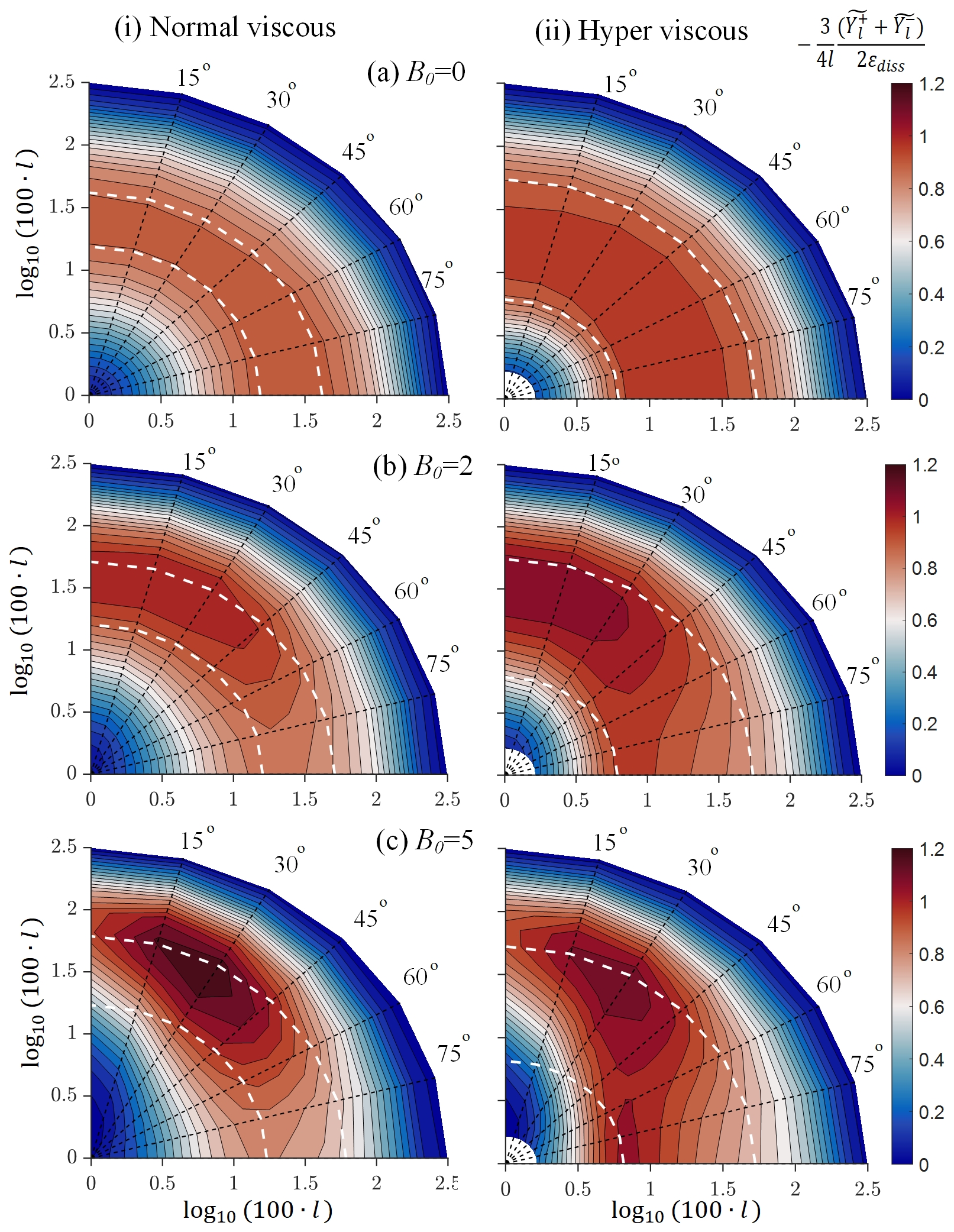}	\\
	\caption{Comparison of normalized third-order structure functions, i.e. $-3(\widetilde{ Y_l^+}+\widetilde{ Y_l^-})/(4\varepsilon_{diss} l) $ between (i) simulations with normal viscosity and (ii) simulations with hyperviscosity with polar angles ranging $\theta=$ [0$^\circ$:15$^\circ$:90$^\circ$]. $\theta=0^\circ$ represents parallel direction relative to $\boldsymbol{B}_0$, and $\theta=90^\circ$ represents perpendicular direction relative to $\boldsymbol{B}_0$. The white dashed lines represent the inertial range, identified with the direction-averaged form of third-order law, where the $-3(\overline{ Y_l^+}+\overline{ Y_l^-})/(4\varepsilon_{diss} l)$ is beyond a threshold, say 0.9. 
	% Dashed lines are guidelines for visualization. %lag $(l)$ is the increment distance in $\delta z$.  
    All contours are integrated over $\phi$ and time. For the visualisation of the inertial range, the $x$-axis is transformed into the $\log_{10}{(100l)}$.}
\label{fig:ComparisonContour_SF_hyper2e7_8e7_V1}
\end{figure} 

\begin{figure}
	\centering
	\includegraphics[width=0.9\linewidth,keepaspectratio]{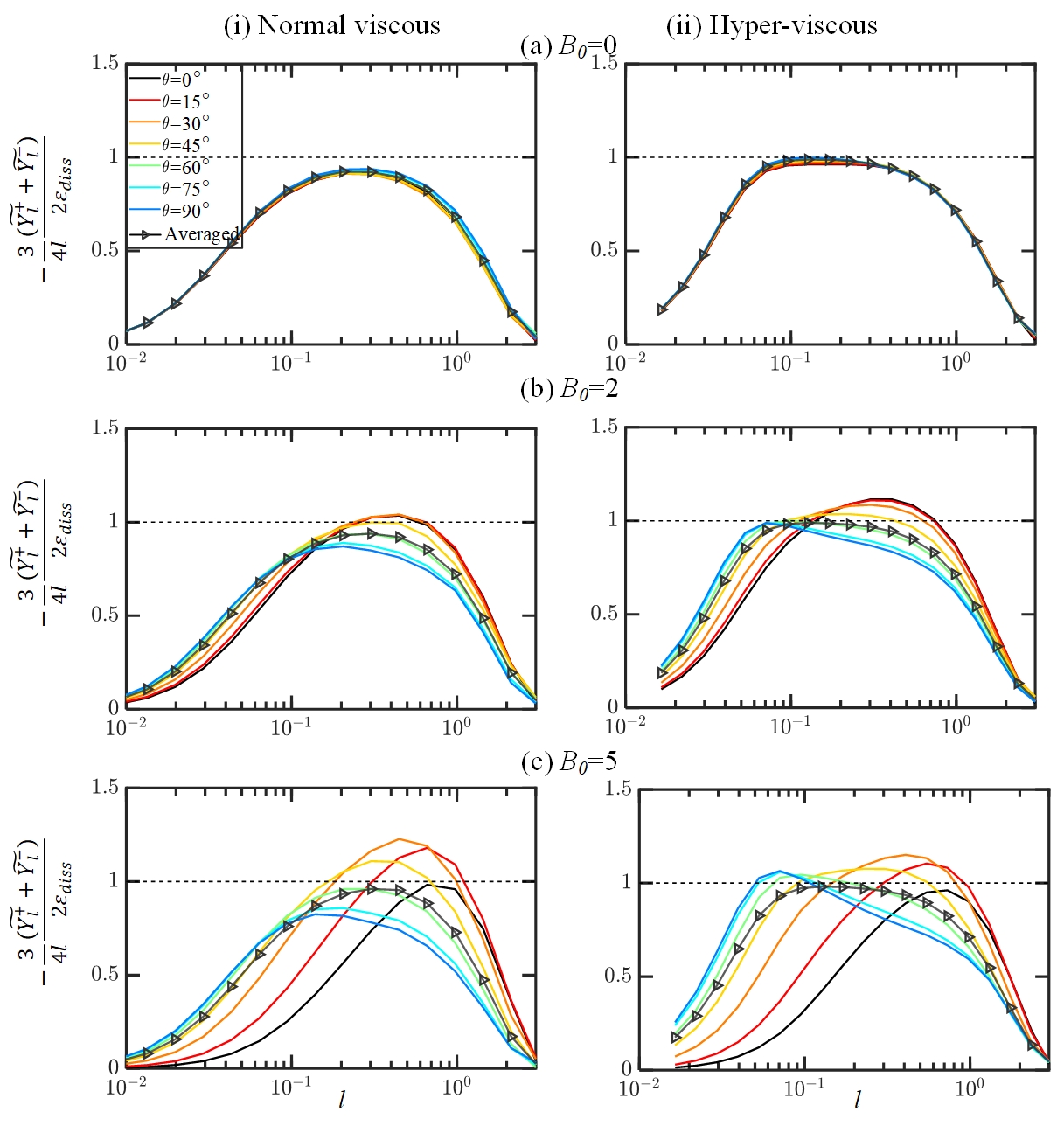}	\\[0.2cm]
 \caption{Comparison of normalized third-order structure functions, i.e. $-3(\widetilde{ Y_l^+}+\widetilde{ Y_l^-})/(4\varepsilon_{diss} l) $, at $B_0=0,2,5$ between (i) normal viscous and (ii) hyper-viscous simulations with $\theta=$ [0$^\circ$:15$^\circ$:90$^\circ$], and the solid line with triangles represents the direction-averaged profile, i.e. $-3(\overline{ Y_l^+}+\overline{ Y_l^-})/(4\varepsilon_{diss} l)$. All curves are averaged over $\phi$ and time. For nonzero $B_0$ the estimated transfer rate peaks at large scales for parallel angles, and at progressively smaller scales for perpendicular angles.
 }
\label{fig:Comparison_SF_hyper2e7_norm8e4}
\end{figure} 
Figure \ref{fig:ComparisonContour_SF_hyper2e7_8e7_V1} shows contour of the averaged and normalized third-order structure functions for both normal viscous and hyper-viscous cases with various values of $B_0$. 
The inertial range are identified with the direction-averaged form of third-order law, where the $-3(\overline{ Y_l^+}+\overline{ Y_l^-})/(4\varepsilon_{diss} l)$ is beyond a threshold, say 0.9, and marked with the white dashed lines. A straightforward observation is that the isotropic cases at $B_0=$ 0 present a distribution of the normalized structure functions that is essentially independent of polar angle $\theta$. Unlike this isotropic case ($B_0$=0), the dark contour lines for $B_0=2$ and 5 do not distribute symmetrically. %deviate from the circular quadrants (white dashed lines).
More specifically, the contour lines at small $l$ (dark blue regions close to origins) are elongated along the parallel direction $\theta=0^\circ$, which can be interpreted as the anisotropy introduced by the mean magnetic field. The contour lines at large $l$ (dark blue outer regions) approach a more circular conformation. This is not in conflict with the frequently observed picture of anisotropy in decaying MHD turbulence, as the present system is driven isotropically at large scales. The inertial range exists in the transition region between small and large scales as marked with the white dashed lines. %Visually, a plateau exists in the middle.
%(which is also called the inertial range), there exists a plateau along each direction, {\color{red}which can be interpreted as the `radial' transfer rate but not rigorously as the energy transfer rate. Since Eq.(\ref{eq:3rd_order_sf_2D_lag}) does not equal the divergence of the structure function and further the energy transfer rate in the anisotropic conditions}. 
For nonzero $B_0$ cases, the peak value of the normalized third-order structure function at larger $\theta$ occurs at smaller scales, which is consistent with the results in Refs. \citep{verdini2015anisotropy,wang2022strategies}. The maximum radial transfer rate shifts away from $B_0$ direction at larger $B_0$, with the corresponding $\theta$ of the maxima for $B_0=$ 2 and 5 is $\theta=0^\circ$ and $\theta=30^\circ$, respectively, which is more clear in Figure \ref{fig:Comparison_SF_hyper2e7_norm8e4}.% This is only true for the case considering the factor of the above-mentioned rate, i.e., the one for the radial rate, but not the contribution, since the latter needs to consider the weights from the angle.
%with the increase of $B_0$, the maximum value of the normalized third-order structure function shifts away from $\theta=0^\circ$. For example, for the normal viscous cases,

We can see the clear difference between normal and hyper-viscous cases in Figure \ref{fig:Comparison_SF_hyper2e7_norm8e4}, where the normalized third-order structure functions at $\theta \in [0^\circ:15^\circ:90^\circ]$ are shown. Also shown is the normalized and direction-averaged third-order structure function, i.e. $-3(\overline{ Y_l^+}+\overline{ Y_l^-})/(4\varepsilon_{diss} l)$. Even though the normalized third-order structure functions exhibit evident dependence on $\theta$, the direction-averaged third-order law attains an accurate cascade rate with less than 5$\%$ error for all cases. The plateau of the direction-averaged form in the Figure \ref{fig:Comparison_SF_hyper2e7_norm8e4} gives a rough idea of the inertial range, which indicates that $l\in[0.15,0.43]$ and $l\in [0.06,0.55]$ at $B_0=0$ can be roughly identified as the inertial range for the normal and hyper-viscous cases, respectively. This elongated inertial range is also consistent with the estimates of the inertial range in Figure \ref{fig:ParPerES_B5rnu2e7_V1}, where $k=\pi/l$. Therefore, as expected, hyperviscosity enables a wider inertial range than normal viscosity and the longer inertial range is beneficial to examine the third-order law. 

The hyper-viscous cases show a similar polar angle dependence to the normal viscous cases at low $B_0$ (i.e., weak anisotropy). At $B_0=0$, the individual $\theta$ profiles in Figure \ref{fig:Comparison_SF_hyper2e7_norm8e4} overlap with the direction-averaged profile, indicating the validity of the 1D isotropic third-order law. As $B_0$ increases, they deviate from the direction-averaged profile. In particular, when $B_0$ is large enough (e.g. the bottom row in Figure \ref{fig:Comparison_SF_hyper2e7_norm8e4} with $B_0=5$), the third-order structure functions for the hyper-viscous cases exhibit distinct peak values from the normal viscous cases. For the normal viscous case, the third-order law tends to underestimate the cascade rate for $\theta>45^\circ$, while overestimate the cascade rate for $\theta<$45$^\circ$. As such, the maximum value of the estimated cascade rate locates at $\theta=30^\circ$, also shown in Figure \ref{fig:ComparisonContour_SF_hyper2e7_8e7_V1}.
In contrast, for the hyper-viscous case, the third-order law at $\theta\sim 90^\circ$ overestimates the cascade rate and the contour map in Figure \ref{fig:ComparisonContour_SF_hyper2e7_8e7_V1} exhibits two local maxima at $\theta=30^\circ$ and $\theta=90^\circ$.
The maximum at $\theta=90^\circ$ could be attributed to the dissipation concentrated at smaller scales due to hyperviscosity. As we can see, the energy transfer changes gradually with angles and the most efficient transfer may not 
necessarily occur in the 
strictly perpendicular direction.

\subsection{Azimuthal angle dependence}

\begin{figure}
	\centering
	\includegraphics[width=0.9\linewidth,keepaspectratio]{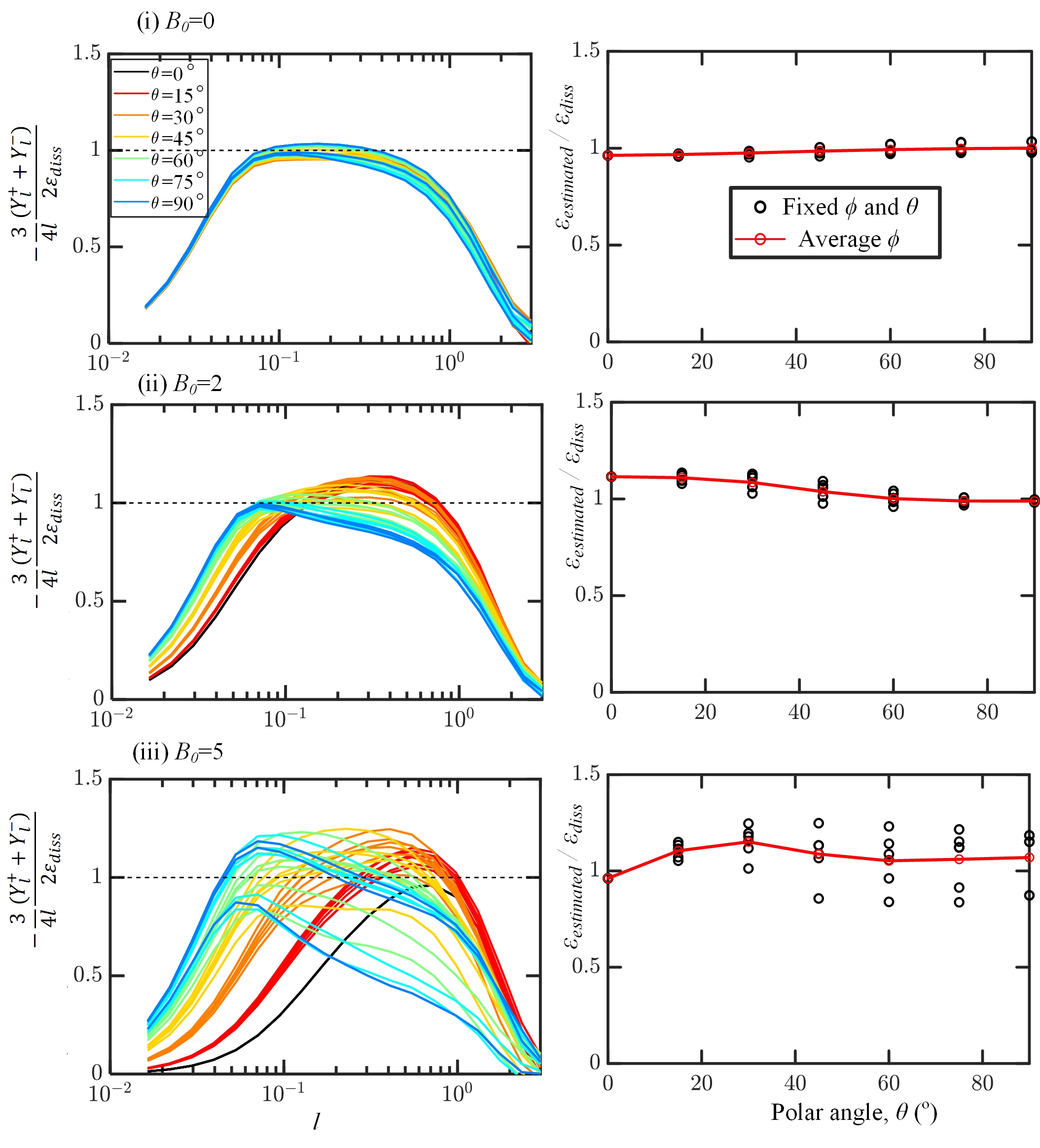}	\\ 
	\caption{(a) Normalized third-order structure functions, i.e. $-3({ Y_l^+}+ Y_l^-)/(4\varepsilon_{diss} l) $, for (i) $B_0=0$, (ii) $B_0=2$ and (iii) $B_0=5$ hyper-viscous cases at different $\theta$ and $\phi$. Each series of colors represents a fixed $\theta$ and varied $\phi$. (b) estimated cascade rates from column (a) by extracting the peaks of each curves and then normalized with the total dissipation rate, $\varepsilon_{diss}$. The solid line with circles represents the averaged profile on the azimuthal direction. All curves are time averaged.} %Bin: should we add the eq. explanation in the caption?
	\label{fig:Singlephi_hyper2e7_B515}
\end{figure}

In this subsection, the hyper-viscous cases will be used to demonstrate the azimuthal dependence of the third-order structure function. 
The left column of Figure \ref{fig:Singlephi_hyper2e7_B515} shows the normalized third-order structure function at different $\theta$ and $\phi$. For the each series of the line with the same color, the $\theta$ is fixed and $\phi$ is varied. It can be seen that the variability of third-order structure functions in the azimuthal and polar angles increases with the increase of $B_0$, indicated by the more scattered distribution of the profiles at higher $B_0$. Specifically, at $B_0=0$, these individual lines almost collapse, indicating the isotropic features in both the polar and azimuthal directions. At $B_0=2$, the peak values of these lines are almost beyond unity. At $B_0=5$, the peaks vary beyond and below the unity. To further quantify these departures from the actual cascade rates in Table \ref{table:setup}, the right column of Figure \ref{fig:Singlephi_hyper2e7_B515} is plotted by using the peak of each profile, representing the estimated cascade rate. The distribution of these estimates, as marked with the dark circles, is more scatter at a larger $B_0$. At $B_0= 2, 5$, the maximum estimated cascade rate can depart from the actual value by 10$\%$ and $25\%$ and both at $\theta=45^\circ$. We expect this departure to be even greater at larger $B_0$. 
%The maximum fluctuations of these peak values for $B_0=2,5$ happen  with 10$\%$ and 40$\%$ maximum variations, respectively. 
From the red line in the right column of Figure \ref{fig:Singlephi_hyper2e7_B515}, after doing the azimuthal average, the maximum departures from unity reduces to 3$\%$ and 15$\%$ at $B_0=2, 5$, respectively. %The $B_0=0,2$ cases do not improve, since its peak value happens at $\theta=0^\circ$ and thus without the azimuthal angle dependence. However, except the parallel directions, the maximum departures from unity reduce from 10$\%$ to 3$\%$ at $B_0=2$, and from 3$\%$ to 0.1$\%$ at $B_0=0$.. the maximum departures from unity reduce

Theoretically, given that the number of frames is enough to do the average, the structure function should be independent of the azimuthal angles, or the energy transfer is isotropic in the perpendicular plane. However, in some contexts, the sampling dataset is limited to finite snapshots as in DNS or directions as in observations, as such the dependence may exist on the azimuthal directions. An adequate coverage of azimuthal angles is required to obtain an accurate estimation of cascade rates. To further check the mutual impact of the time and angle averages, the effects of the sampling time will be discussed in the next subsection. % highly aligned with LC's point with a discussion in the end of the section or a conclusive words. It is widely accepted that the dynamics in the 2D plane perpendicular to the mean magnetic field may be regarded as isotropic (i.e., `gyrotropic'). A significant point found here is the recognition that in MHD turbulence with strong mean magnetic field, {\color{red}the energy transfer is not isotropic in the perpendicular plane; nor is it azimuthally independent. } % Also, this former sentence should combine with the last paragraph in this paragraph.

\subsection{Effects of time averaging}
In this subsection, the hyper-viscous cases will be used to demonstrate the effect of time averaging. At least two aspects of the time averaging are of central importance, one being the time interval (or sampling frequency) $\Delta T$, and the other being the length of periods $T$. We can expect that in our driven cases, the smaller $\Delta T$ and the longer $T$, the more reliable statistics. The time interval $\Delta T$ ($\Delta T < T_e$, where $T_e$ is the large-eddy turnover time) does not show significant impacts on the third-order structure function in our cases (Figures are not shown here). Therefore, in the following analysis, we fix $\Delta T=0.5T_e$ and focus on the effect of the length of periods $T$. These periods are within the statistically stationary periods listed in Table \ref{table:setup}. % as demonstrated by the total energy and dissipation rate evolution in Appendix A. % should we consider to put the time correlation here?
 
% YYang suggest to delete B=15 case in Fig.3.4
\begin{figure}
	\centering
	\includegraphics[width=1\linewidth,keepaspectratio]{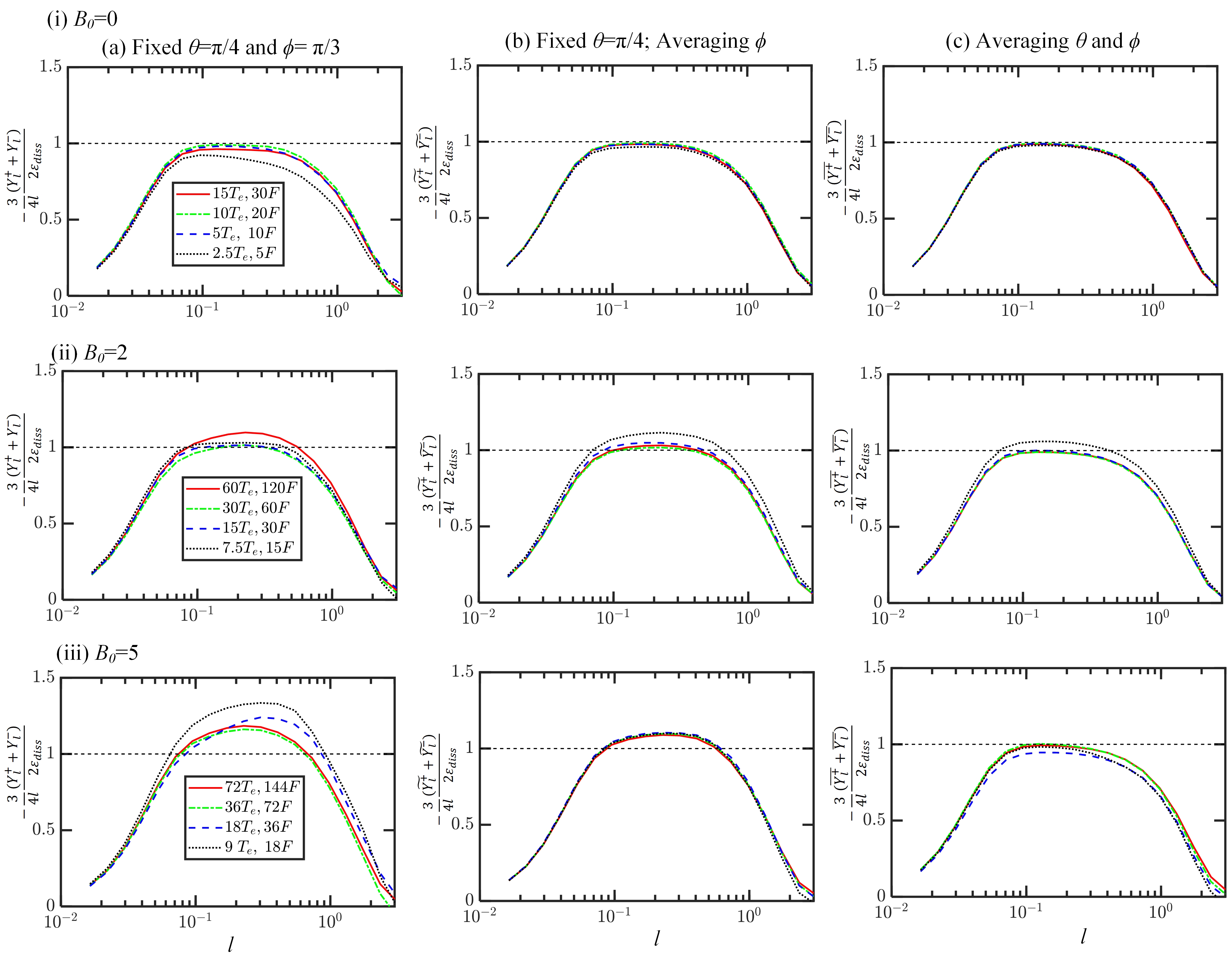}	\\ 
	\caption{Effects of the number of time frames on the calculation of the normalized third-order structure functions  at (i) $B_0=0$, (ii) $B_0=2$  and (iii) $B_0=5$ for the hyper-viscous cases. Left column (a): fixed $\theta = \pi/4; \phi = \pi/3$. Middle column (b): fixed $\theta = \pi/4$ and azimuthal-averaged profiles.  Right column (c): direction-averaged profiles. $T_e$ denotes the large-eddy turnover time at $B_0=0$, and $F$ represents the number of frames  used in the time average.}
	\label{fig:AVGtime_hyper2e7_B515}
\end{figure}
%Considering that a big time interval reduces the time correlation between two instants, there is no maximum sampling time interval. One should not choose too big time interval to save the computational cost, given that the total number of frames should be enough to do average. The present work used 0.5 $T_e$ time interval for all the hyper-viscous simulations. This value is a compromise between the sampling frequency and the total number of frames to attain the converged structure function profiles in the whole sampling period. This period should be statistically stationary without statistical bias as demonstrated by the total energy and dissipation rate evolution in the Appendix A. This is also verified by structure function profiles with fixed period length, not shown here for brevity.

Figure \ref{fig:AVGtime_hyper2e7_B515} shows the effect of the length of periods (or the number of time frames, $F$) on the third-order structure function. The third-order law at fixed $\theta$ and $\phi$ as Eq. (\ref{eq:3rd_order_sf_3D_lag}), at fixed $\theta$ but $\phi$ averaged as Eq. (\ref{eq:3rd_order_sf_2D_lag}) and both $\theta$ and $\phi$ averaged as Eq. (\ref{eq:3rd_order_sf_1D_lag}) is shown in left, middle and right columns, respectively. Three key observations can be found: % this eq. index is wrong.  

i) the direction-averaged profiles require less time frames ($F$)
to converge than the profiles with fixed $\theta$ and $\phi$. For instance, at $B_0=$ 5 the direction-averaged profiles converge with 36 time snapshots, while the profiles with fixed $\theta$ and $\phi$ are not converged until  72 time frames are employed.  

ii) a smaller $B_0$ requires  a lesser number  of frames to converge. The required number of time frames for $B_0=$ 0 and $B_0=$ 5 to converge are 
about 10 and 70,
respectively, without the directional average (column (a) in Figure \ref{fig:AVGtime_hyper2e7_B515}).
These times 
are reduced to about 5 and 30, respectively, with 
directional averaging (see column (c) in Figure \ref{fig:AVGtime_hyper2e7_B515}.)

iii) sufficient time averaging cannot make up for the lack of angle coverage, indicated by the different peak values of non-direction-averaged and direction-averaged profiles in the anisotropic cases. Also, by comparing the three columns, the angle averaging makes the plateaus of these plots closer to the actual dissipation rate. For instance, at $B_0=5$, the converged profiles without angular averaging, with only azimuthal averaging and with both azimuthal and polar averaging attain peak values of 1.19, 1.09 and 0.98, respectively. This indicates that, compared with the time averaging, the angle averaging is a more effective way to make the calculation of the structure function converge.

\section{Conclusions}
\label{sec:conclusions}
3D simulations of incompressible MHD turbulence with normal and hyperviscosity were conducted with different externally supported (mean) uniform magnetic fields to systematically study the effect of external mean magnetic fields on the cross-scale energy transfer and the third-order law. Three different forms of third-order structure functions were calculated with or without the averaging the azimuthal and polar angles. The results show that, compared with the normal viscous cases, the hyperviscosity elongates and separates the inertial range from the dissipation range, thus helping the examination of the above-mentioned forms of third-order law. % with much lower computational cost.  and  
%Such an extended inertial range leads to enhance the energy transfer in the small scales. Specifically, the azimuthally averaged third-order structure function in the anisotropic hyperviscous case show a higher peak value in perpendicular to $B_0$ direction than the normal-viscous case.
% shows a , located at the connection between the inertial and dissipation ranges. This is perhaps due to the bottleneck effect, and caution is needed in related studies on the dissipation range. 

The direction-averaged third-order law can predict the energy cascade rates and inertial range accurately, even at very high $B_0$ (see 
background on this in \cite{TaylorEA03} and \cite{wang2022strategies}). However, the third-order law is highly dependent on the polar angle in anisotropic MHD turbulence. The azimuthal angle dependence we find here is counterintuitive, in that the statistics are usually deemed to be isotropic in the perpendicular plane to $B_0$. This azimuthal angle dependence was further confirmed with an examination of the number of averaging time frames. 
More time frames are required to make the calculation of the third-order structure function converge in the anisotropic condition, especially at high $B_0$. However, even using the enough time frames, the structure function profiles with and without the azimuthal average are still different. We conclude that the time averaging cannot make up for a lack of angle-averaging, especially at large values of mean magnetic field $B_0$. %{\color{red}This azimuthal dependence may be from the periodic boundary or the domain size, and further investigation is deserved.} %, {\color{red}provided that the time frames is enough to do the average.}

% {\color{red}Another interesting observation for the anisotropic cases, especially in the hyperviscous cases, is that there is no obvious plateau in the perpendicular direction of the normalized and azimuthal-averaged third-order structure function, corresponding to the inertial range marked with the direction-averaged third-order law. A potential scaling exists in this range and deserves to future study. Also, it may relate to the contribution of the polar transfer rate to the total transfer rate at different scales. In other words, the transfer rate is enhanced in the smaller scale as closer to the perpendicular plane.}

An interesting finding, also seen in earlier studies \cite{VerdiniEA15,wang2022strategies}, is that identified peaks or plateaus of the estimated transfer rate occur at different scales for different angles to the mean magnetic field. Accordingly, this also means that the inertial range will be assigned to different ranges of scales. Our use of hyperviscosity shows that this method produces an extended plateau, facilitating analysis of inertial range properties. Future studies may find a rationale for understanding how these different ranges come into being, as the balance of local transfer rate competes with dissipation at different angles. Such analyses may require examination of the {\it vector} character of the Yaglom energy flux, a subject that we have not engaged in here. It should be noted however that there are some studies that have begun such analyses, typically by making assumptions regarding the symmetry of the transfer. Examples include the 2D (perpendicular) + 1D (parallel) model employed by \cite{MacBride08,Stawarz09,StawarzEA10}. Such models may require further refinement since at least some prior models enforce the questionable assumption that the transfers of energy in parallel and perpendicular directions are independent.\\
% Just a final check, is the citation on the first sentence of this paragraph correct for Yang's paper or Wang's paper. I believe that it should be Wang's paper, since Yang's paper does not look into the angle effect. Am I right? 

% The total sampling frames are about 30 for the direction-averaged plots and doubled for the profiles with fixed polar and azimuthal angles.

%% Acknowledgements
% \section{Acknowledgements}
% \label{sec:ack}

\noindent \textbf{Funding.} \; We acknowledge the support from NSFC (Grant Nos. 12225204, and 11902138); Department of Science and Technology of Guangdong Province (Grant Nos. 2019B21203001 and 2020B1212030001); the Shenzhen Science and Technology Program (Grant No. KQTD20180411143441009). Y.Y. and W.H.M. are supported by NSF grant AGS02108834 and by NASA under the IMAP project at Princeton (subcontract 0000317 to Delaware). W.H.M.is also supported as a Co-Investigator on the Helioswarm project.
The assistance of resources and services from the Center for Computational Science and Engineering of Southern University of Science and Technology is also acknowledged. \\
% Bin: to be confirmed.
% National Natural Science Foundation of China, grant no. 11902138, 91752201.

\noindent \textbf{Declaration of interests.} The authors report no conflict of interest.

\appendix

\bibliographystyle{jfm}
\bibliography{3rdLaw23}

\newcommand{\BIBand} {and} %...... how 'and' appears in authors
  \newcommand{\boldVol}[1] {\textbf{#1}} %......................
  \providecommand{\SortNoop}[1]{} %.......Use as {\SortNoop{Aaa}}
  \providecommand{\sortnoop}[1]{} %..............................
  \newcommand{\stereo} {\emph{{S}{T}{E}{R}{E}{O}}} %.................
  \newcommand{\au} {{A}{U}\ } %..................................
  \newcommand{\AU} {{A}{U}\ } %.................................
  \newcommand{\MHD} {{M}{H}{D}\ } %..............................
  \newcommand{\mhd} {{M}{H}{D}\ } %...............................
  \newcommand{\RMHD} {{R}{M}{H}{D}\ } %...........................
  \newcommand{\rmhd} {{R}{M}{H}{D}\ } %...........................
  \newcommand{\wkb} {{W}{K}{B}\ } %..............................
  \newcommand{\alfven} {{A}lfv{\'e}n\ } %...........................
  \newcommand{\alfvenic} {{A}lfv{\'e}nic\ } %.........................
  \newcommand{\Alfven} {{A}lfv{\'e}n\ } %...........................
  \newcommand{\Alfvenic} {{A}lfv{\'e}nic\ }
\begin{thebibliography}{61}
\expandafter\ifx\csname natexlab\endcsname\relax\def\natexlab#1{#1}\fi
\def\au#1{#1} \def\ed#1{#1} \def\yr#1{#1}\def\at#1{#1}\def\jt#1{\textit{#1}}
  \def\bt#1{#1}\def\bvol#1{\textbf{#1}} \def\vol#1{#1} \def\pg#1{#1}
  \def\publ#1{#1}\def\arxiv#1{#1}\def\org#1{#1}\def\st#1{\textit{#1}}

\bibitem[Alexakis {\em et~al.\/}(2007{\natexlab{{\em a\/}}})Alexakis, Bigot,
  Politano \& Galtier]{Alexakis07a}
{\sc \au{Alexakis, A.}, \au{Bigot, B.}, \au{Politano, H.} \& \au{Galtier, S.}}
  \yr{2007{\natexlab{{\em a\/}}}}  \at{Anisotropic fluxes and nonlocal
  interactions in magnetohydrodynamic turbulence}.  \jt{Phys.~Rev.~E}
  \bvol{76},  \pg{056313}.

\bibitem[Alexakis {\em et~al.\/}(2007{\natexlab{{\em b\/}}})Alexakis, Bigot,
  Politano \& Galtier]{alexakis2007anisotropic}
{\sc \au{Alexakis, A}, \au{Bigot, B}, \au{Politano, H{\'e}l{\`e}ne} \&
  \au{Galtier, S}} \yr{2007{\natexlab{{\em b\/}}}}  \at{Anisotropic fluxes and
  nonlocal interactions in magnetohydrodynamic turbulence}.  \jt{Physical
  Review E}  \bvol{76}~(5),  \pg{056313}.

\bibitem[Andr\'es {\em et~al.\/}(2019)Andr\'es, Sahraoui, Galtier, Hadid,
  Ferrand \& Huang]{AndresEA19}
{\sc \au{Andr\'es, N.}, \au{Sahraoui, F.}, \au{Galtier, S.}, \au{Hadid, L.~Z.},
  \au{Ferrand, R.} \& \au{Huang, S.~Y.}} \yr{2019}  \at{Energy cascade rate
  measured in a collisionless space plasma with mms data and compressible hall
  magnetohydrodynamic turbulence theory}.  \jt{Phys. Rev. Lett.}  \bvol{123},
  \pg{245101}.

\bibitem[Antonia {\em et~al.\/}(1997)Antonia, Ould-Rouis, Anselmet \&
  Zhu]{antonia1997analogy}
{\sc \au{Antonia, RA}, \au{Ould-Rouis, M}, \au{Anselmet, F} \& \au{Zhu, Y}}
  \yr{1997}  \at{Analogy between predictions of kolmogorov and yaglom}.
  \jt{Journal of Fluid Mechanics}  \bvol{332},  \pg{395--409}.

\bibitem[{Bandyopadhyay} {\em et~al.\/}(2020){Bandyopadhyay}, {Sorriso-Valvo},
  {Chasapis}, {Hellinger}, {Matthaeus}, {Verdini}, {Landi}, {Franci},
  {Matteini}, {Giles}, {Gershman}, {Moore}, {Pollock}, {Russell}, {Strangeway},
  {Torbert} \& {Burch}]{BandyopadhyayEA20-Hall}
{\sc \au{{Bandyopadhyay}, Riddhi}, \au{{Sorriso-Valvo}, Luca}, \au{{Chasapis},
  Alexand~ros}, \au{{Hellinger}, Petr}, \au{{Matthaeus}, William~H.},
  \au{{Verdini}, Andrea}, \au{{Landi}, Simone}, \au{{Franci}, Luca},
  \au{{Matteini}, Lorenzo}, \au{{Giles}, Barbara~L.}, \au{{Gershman},
  Daniel~J.}, \au{{Moore}, Thomas~E.}, \au{{Pollock}, Craig~J.}, \au{{Russell},
  Christopher~T.}, \au{{Strangeway}, Robert~J.}, \au{{Torbert}, Roy~B.} \&
  \au{{Burch}, James~L.}} \yr{2020}  \at{{In Situ Observation of Hall
  Magnetohydrodynamic Cascade in Space Plasma}}.  \jt{Phys Rev. Lett.}
  \bvol{124}~(22),  \pg{225101},  \arxiv{arXiv: 1907.06802}.

\bibitem[Banerjee {\em et~al.\/}(2016)Banerjee, Hadid, Sahraoui \&
  Galtier]{Banerjee16b}
{\sc \au{Banerjee, S.}, \au{Hadid, L.~Z.}, \au{Sahraoui, F.} \& \au{Galtier,
  S.}} \yr{2016}  \at{Scaling of compressible magnetohydrodynamic turbulence in
  the fast solar wind}.  \jt{Astrophys.~J.~Lett.}  \bvol{829},  \pg{L27}.

\bibitem[Beresnyak \& Lazarian(2009)]{beresnyak2009comparison}
{\sc \au{Beresnyak, Andrey} \& \au{Lazarian, Alex}} \yr{2009}  \at{Comparison
  of spectral slopes of magnetohydrodynamic and hydrodynamic turbulence and
  measurements of alignment effects}.  \jt{The Astrophysical Journal}
  \bvol{702}~(2),  \pg{1190}.

\bibitem[Biskamp(2003)]{biskamp2003magnetohydrodynamic}
{\sc \au{Biskamp, Dieter}} \yr{2003} {\em Magnetohydrodynamic turbulence\/}.
  \publ{Cambridge University Press}.

\bibitem[Biskamp \& M{\"u}ller(2000)]{biskamp2000scaling}
{\sc \au{Biskamp, Dieter} \& \au{M{\"u}ller, Wolf-Christian}} \yr{2000}
  \at{Scaling properties of three-dimensional isotropic magnetohydrodynamic
  turbulence}.  \jt{Physics of Plasmas}  \bvol{7}~(12),  \pg{4889--4900}.

\bibitem[{Bruno} \& {Carbone}(2013)]{BrunoCarboneLRSP13}
{\sc \au{{Bruno}, Roberto} \& \au{{Carbone}, Vincenzo}} \yr{2013}  \at{{The
  Solar Wind as a Turbulence Laboratory}}.  \jt{Living Reviews in Solar
  Physics}  \bvol{10}~(1),  \pg{2}.

\bibitem[Carbone {\em et~al.\/}(2009)Carbone, Marino, Sorriso-Valvo, Noullez \&
  Bruno]{Carbone09}
{\sc \au{Carbone, V.}, \au{Marino, R.}, \au{Sorriso-Valvo, L.}, \au{Noullez,
  A.} \& \au{Bruno, R.}} \yr{2009}  \at{Scaling laws of turbulence and heating
  of fast solar wind: the role of density fluctuations}.  \jt{Phys.~Rev.~Lett.}
   \bvol{103},  \pg{061102}.

\bibitem[Coleman(1968)]{Coleman68}
{\sc \au{Coleman, P.~J.}} \yr{1968}  \at{Turbulence, viscosity, and dissipation
  in the solar wind plasma}.  \jt{Astrophys.\ J.}  \bvol{153},  \pg{371--388}.

\bibitem[De~Karman \& Howarth(1938)]{de1938statistical}
{\sc \au{De~Karman, Theodore} \& \au{Howarth, Leslie}} \yr{1938}  \at{On the
  statistical theory of isotropic turbulence}.  \jt{Proceedings of the Royal
  Society of London. Series A-Mathematical and Physical Sciences}
  \bvol{164}~(917),  \pg{192--215}.

\bibitem[Ferrand {\em et~al.\/}(2019)Ferrand, Galtier, Sahraoui, Meyrand,
  Andr{\'e}s \& Banerjee]{Ferrand2019exact}
{\sc \au{Ferrand, Renaud}, \au{Galtier, S{\'e}bastien}, \au{Sahraoui, Fouad},
  \au{Meyrand, Romain}, \au{Andr{\'e}s, Nahuel} \& \au{Banerjee, Supratik}}
  \yr{2019}  \at{On exact laws in incompressible hall magnetohydrodynamic
  turbulence}.  \jt{The Astrophysical Journal}  \bvol{881}~(1),  \pg{50}.

\bibitem[Ferrand {\em et~al.\/}(2022)Ferrand, Sahraoui, Galtier, Andr{\'e}s,
  Mininni \& Dmitruk]{ferrand2022depth}
{\sc \au{Ferrand, R}, \au{Sahraoui, F}, \au{Galtier, S}, \au{Andr{\'e}s, N},
  \au{Mininni, P} \& \au{Dmitruk, P}} \yr{2022}  \at{An in-depth numerical
  study of exact laws for compressible hall magnetohydrodynamic turbulence}.
  \jt{arXiv preprint arXiv:2201.10819} .

\bibitem[{Frisch}(1995)]{FrischBook95}
{\sc \au{{Frisch}, U.}} \yr{1995} {\em {Turbulence. The legacy of A. N.
  Kolmogorov.}\/}.

\bibitem[Frish(1995)]{frish1995turbulence}
{\sc \au{Frish, U}} \yr{1995}  \at{Turbulence: The legacy of an kolmogorov
  cambridge}.  \jt{New York} .

\bibitem[Galtier(2009)]{galtier2009exact}
{\sc \au{Galtier, S}} \yr{2009}  \at{Exact vectorial law for axisymmetric
  magnetohydrodynamics turbulence}.  \jt{The Astrophysical Journal}
  \bvol{704}~(2),  \pg{1371}.

\bibitem[Gogoberidze {\em et~al.\/}(2013)Gogoberidze, Perri \&
  Carbone]{Gogoberidze2013yaglom}
{\sc \au{Gogoberidze, G}, \au{Perri, S} \& \au{Carbone, V}} \yr{2013}  \at{The
  yaglom law in the expanding solar wind}.  \jt{The Astrophysical Journal}
  \bvol{769}~(2),  \pg{111}.

\bibitem[Hadid {\em et~al.\/}(2017)Hadid, Sahraoui \& Galtier]{Hadid2017energy}
{\sc \au{Hadid, LZ}, \au{Sahraoui, F} \& \au{Galtier, S}} \yr{2017}  \at{Energy
  cascade rate in compressible fast and slow solar wind turbulence}.  \jt{The
  Astrophysical Journal}  \bvol{838}~(1),  \pg{9}.

\bibitem[Hellinger {\em et~al.\/}(2013)Hellinger, Tr{\'a}vn{\'\i}{\v{c}}ek,
  {\v{S}}tver{\'a}k, Matteini \& Velli]{Hellinger2013proton}
{\sc \au{Hellinger, Petr}, \au{Tr{\'a}vn{\'\i}{\v{c}}ek, Pavel~M},
  \au{{\v{S}}tver{\'a}k, {\v{S}}t{\v{e}}p{\'a}n}, \au{Matteini, Lorenzo} \&
  \au{Velli, Marco}} \yr{2013}  \at{Proton thermal energetics in the solar
  wind: Helios reloaded}.  \jt{Journal of Geophysical Research: Space Physics}
  \bvol{118}~(4),  \pg{1351--1365}.

\bibitem[Hellinger {\em et~al.\/}(2018)Hellinger, Verdini, Landi, Franci \&
  Matteini]{Hellinger2018karman}
{\sc \au{Hellinger, Petr}, \au{Verdini, Andrea}, \au{Landi, Simone},
  \au{Franci, Luca} \& \au{Matteini, Lorenzo}} \yr{2018}  \at{von
  k{\'a}rm{\'a}n--howarth equation for hall magnetohydrodynamics: Hybrid
  simulations}.  \jt{The Astrophysical Journal Letters}  \bvol{857}~(2),
  \pg{L19}.

\bibitem[Horbury {\em et~al.\/}(2008)Horbury, Forman \& Oughton]{HorburyEA08}
{\sc \au{Horbury, T.~S.}, \au{Forman, M.} \& \au{Oughton, S.}} \yr{2008}
  \at{Anisotropic scaling of magnetohydrodynamic turbulence}.  \jt{Phys. Rev.
  Lett.}  \bvol{101}.

\bibitem[Hossain {\em et~al.\/}(1995)Hossain, Gray, Pontius~Jr, Matthaeus \&
  Oughton]{hossain1995phenomenology}
{\sc \au{Hossain, Murshed}, \au{Gray, Perry~C}, \au{Pontius~Jr, Duane~H},
  \au{Matthaeus, William~H} \& \au{Oughton, Sean}} \yr{1995}  \at{Phenomenology
  for the decay of energy-containing eddies in homogeneous mhd turbulence}.
  \jt{Physics of Fluids}  \bvol{7}~(11),  \pg{2886--2904}.

\bibitem[{Jokipii} \& {Hollweg}(1970)]{JokipiiHollweg70}
{\sc \au{{Jokipii}, J.~R.} \& \au{{Hollweg}, J.~V.}} \yr{1970}
  \at{{Interplanetary Scintillations and the Structure of Solar-Wind
  Fluctuations}}.  \jt{apj}  \bvol{160},  \pg{745}.

\bibitem[de~K{\'a}rm{\'a}n \& Howarth(1938)]{KarmanHowarth38}
{\sc \au{de~K{\'a}rm{\'a}n, T.} \& \au{Howarth, L.}} \yr{1938}  \at{On the
  statistical theory of isotropic turbulence}.  \jt{Proc. Roy. Soc. London Ser.
  A}  \bvol{164},  \pg{192--215}.

\bibitem[Kolmogorov(1941{\natexlab{{\em a\/}}})]{kolmogorov1941dissipation}
{\sc \au{Kolmogorov, AN}} \yr{1941{\natexlab{{\em a\/}}}} Dissipation of energy
  in locally isotropic turbulence in an incompressible viscous liquid.  \bt{In
  {\em Dokl. Akad. Nauk SSSR\/}}, ,  \vol{vol.~30},  \pg{pp. 299--303}.

\bibitem[Kolmogorov(1941{\natexlab{{\em b\/}}})]{Kol41a}
{\sc \au{Kolmogorov, A.~N.}} \yr{1941{\natexlab{{\em b\/}}}}
  \at{\sortnoop{1941a}{Local} structure of turbulence in an incompressible
  viscous fluid at very high {Reynolds} numbers}.  \jt{Dokl. Akad. Nauk SSSR}
  \bvol{30},  \pg{301--305}, [Reprinted in Proc.\ R.\ Soc.\ London, Ser.\ A
  \textbf{434}, 9--13 (1991)].

\bibitem[Kolmogorov(1941{\natexlab{{\em c\/}}})]{Kol41b}
{\sc \au{Kolmogorov, A.~N.}} \yr{1941{\natexlab{{\em c\/}}}}
  \at{\sortnoop{1941b}{On} degeneration of isotropic turbulence in an
  incompressible viscous liquid}.  \jt{C.R. Acad. Sci. U.R.S.S.}  \bvol{31},
  \pg{538--540}.

\bibitem[Kraichnan(1971)]{Kraichnan71-jfm}
{\sc \au{Kraichnan, R.~H.}} \yr{1971}  \at{Inertial-range transfer in two- and
  three-dimensional turbulence}.  \jt{J.\ Fluid Mech.}  \bvol{47},  \pg{525}.

\bibitem[Kritsuk {\em et~al.\/}(2009)Kritsuk, Ustyugov, Norman \&
  Padoan]{Kritsuk09}
{\sc \au{Kritsuk, A.~G.}, \au{Ustyugov, S.~D.}, \au{Norman, M.~K.} \&
  \au{Padoan, P.}} \yr{2009}  \at{Simulating supersonic turbulence in
  magnetized molecular clouds}.  \jt{J.~Phys.~Conf.~Ser.}  \bvol{180},
  \pg{012020}.

\bibitem[MacBride \& Smith(2008)]{MacBride08}
{\sc \au{MacBride, Benjamin~T.} \& \au{Smith, Charles~W.}} \yr{2008}  \at{The
  turbulent cascade at 1 {AU}: energy transfer and the third-order scaling for
  {MHD}}.  \jt{Astrophys.~J.}  \bvol{679},  \pg{1644--1660}.

\bibitem[Matthaeus {\em et~al.\/}(2019)Matthaeus, Bandyopadhyay, Brown,
  Borovsky, Carbone, Caprioli, Chasapis, Chhiber, Dasso, Dmitruk, Del~Zanna,
  Dmitruk, Franci, Gary, Goldstein, Gomez, Greco, Horbury, Ji, Kasper, Klein,
  Landi, Li, Malara, Maruca, Mininni, Oughton, Papini, Parashar, Petrosyan,
  Pouquet, Retino, Roberts, Ruffolo, Servidio, Spence, Smith, Stawarz,
  TenBarge, Vasquez1, Vaivads, Valentini, Velli, Verdini, Verscharen,
  Whittlesey, Wicks, Bruno \& Zimbardo]{MatthaeusEA19-whitepaper}
{\sc \au{Matthaeus, W.~H.}, \au{Bandyopadhyay, R.}, \au{Brown, M.~R.},
  \au{Borovsky, J.}, \au{Carbone, V.}, \au{Caprioli, D.}, \au{Chasapis, A.},
  \au{Chhiber, R.}, \au{Dasso, S.}, \au{Dmitruk, P.}, \au{Del~Zanna, L.},
  \au{Dmitruk, P.~A.}, \au{Franci, Luca}, \au{Gary, S.~P.}, \au{Goldstein,
  M.~L.}, \au{Gomez, D.}, \au{Greco, A.}, \au{Horbury, T.~S.}, \au{Ji, Hantao},
  \au{Kasper, J.~C.}, \au{Klein, K.~G.}, \au{Landi, S.}, \au{Li, Hui},
  \au{Malara, F.}, \au{Maruca, B.~A.}, \au{Mininni, P.}, \au{Oughton, Sean},
  \au{Papini, E.}, \au{Parashar, T.~N.}, \au{Petrosyan, Arakel}, \au{Pouquet,
  Annick}, \au{Retino, A.}, \au{Roberts, Owen}, \au{Ruffolo, David},
  \au{Servidio, Sergio}, \au{Spence, Harlan}, \au{Smith, C.~W.}, \au{Stawarz,
  J.~E.}, \au{TenBarge, Jason}, \au{Vasquez1, B.~J.}, \au{Vaivads, Andris},
  \au{Valentini, F.}, \au{Velli, Marco}, \au{Verdini, A.}, \au{Verscharen,
  Daniel}, \au{Whittlesey, Phyllis}, \au{Wicks, Robert}, \au{Bruno, R.} \&
  \au{Zimbardo, G.}} \yr{2019} [plasma 2020 decadal] the essential role of
  multi-point measurements in turbulence investigations: the solar wind beyond
  single scale and beyond the taylor hypothesis.

\bibitem[Matthaeus {\em et~al.\/}(1996)Matthaeus, Ghosh, Oughton \&
  Roberts]{MattEA96-var}
{\sc \au{Matthaeus, W.~H.}, \au{Ghosh, S.}, \au{Oughton, S.} \& \au{Roberts,
  D.~A.}} \yr{1996}  \at{Anisotropic three-dimensional \mhd turbulence}.
  \jt{J.\ Geophys.\ Res.}  \bvol{101},  \pg{7619--7629}.

\bibitem[Matthaeus \& Goldstein(1982)]{MattGold82a}
{\sc \au{Matthaeus, W.~H.} \& \au{Goldstein, M.~L.}} \yr{1982}  \at{Measurement
  of the rugged invariants of magnetohydrodynamic turbulence in the solar
  wind}.  \jt{J.\ Geophys.\ Res.}  \bvol{87},  \pg{6011--6028}.

\bibitem[Monin \& Yaglom(1975)]{monin1975statistical}
{\sc \au{Monin, AS} \& \au{Yaglom, AM}} \yr{1975} Statistical fluid mechanics:
  Mechanics of turbulence, vol. 2, 874 pp.

\bibitem[Nie \& Tanveer(1999)]{NieTanveer99}
{\sc \au{Nie, Qing} \& \au{Tanveer, S}} \yr{1999}  \at{A note on third--order
  structure functions in turbulence}.  \jt{Proceedings of the Royal Society of
  London. Series A: Mathematical, Physical and Engineering Sciences}
  \bvol{455}~(1985),  \pg{1615--1635}.

\bibitem[Osman {\em et~al.\/}(2011{\natexlab{{\em a\/}}})Osman, Matthaeus,
  Greco \& Servidio]{Osman11-ApJ}
{\sc \au{Osman, K.~T.}, \au{Matthaeus, W.~H.}, \au{Greco, A.} \& \au{Servidio,
  S.}} \yr{2011{\natexlab{{\em a\/}}}}  \at{Evidence for inhomogeneous heating
  in the solar wind}.  \jt{Astrophys.~J.~Lett.}  \bvol{727},  \pg{L11}.

\bibitem[Osman {\em et~al.\/}(2011{\natexlab{{\em b\/}}})Osman, Wan, Matthaeus,
  Weygand \& Dasso]{Osman11-PRL}
{\sc \au{Osman, K.~T.}, \au{Wan, M.}, \au{Matthaeus, W.~H.}, \au{Weygand,
  J.~M.} \& \au{Dasso, S.}} \yr{2011{\natexlab{{\em b\/}}}}  \at{Anisotropic
  third-moment estimates of the energy cascade in solar wind turbulence using
  multispacecraft data}.  \jt{Phys.~Rev.~Lett.}  \bvol{107},  \pg{165001}.

\bibitem[{Oughton} {\em et~al.\/}(2015){Oughton}, {Matthaeus}, {Wan} \&
  {Osman}]{OughtonEA15_anisotropy}
{\sc \au{{Oughton}, S.}, \au{{Matthaeus}, W.~H.}, \au{{Wan}, M.} \&
  \au{{Osman}, K.~T.}} \yr{2015}  \at{{Anisotropy in solar wind plasma
  turbulence}}.  \jt{Phil Trans. Roy. Soc. A}  \bvol{373}.

\bibitem[Parker(1979)]{Parker-cmf}
{\sc \au{Parker, E.~N.}} \yr{1979} {\em Cosmical Magnetic Fields: {Their}
  Origin and Activity\/}.  \publ{Oxford, UK: Oxford Univeristy Press}.

\bibitem[Podesta(2008)]{Podesta2008laws}
{\sc \au{Podesta, JJ}} \yr{2008}  \at{Laws for third-order moments in
  homogeneous anisotropic incompressible magnetohydrodynamic turbulence}.
  \jt{Journal of Fluid Mechanics}  \bvol{609},  \pg{171--194}.

\bibitem[Podesta {\em et~al.\/}(2007)Podesta, Forman \&
  Smith]{podesta2007anisotropic}
{\sc \au{Podesta, JJ}, \au{Forman, MA} \& \au{Smith, CW}} \yr{2007}
  \at{Anisotropic form of third-order moments and relationship to the cascade
  rate in axisymmetric magnetohydrodynamic turbulence}.  \jt{Physics of
  Plasmas}  \bvol{14}~(9),  \pg{092305}.

\bibitem[Politano \& Pouquet(1998)]{politano1998karman}
{\sc \au{Politano, H} \& \au{Pouquet, A}} \yr{1998}  \at{von
  k{\'a}rm{\'a}n--howarth equation for magnetohydrodynamics and its
  consequences on third-order longitudinal structure and correlation
  functions}.  \jt{Physical Review E}  \bvol{57}~(1),  \pg{R21}.

\bibitem[Shebalin {\em et~al.\/}(1983)Shebalin, Matthaeus \&
  Montgomery]{shebalin1983anisotropy}
{\sc \au{Shebalin, John~V}, \au{Matthaeus, William~H} \& \au{Montgomery,
  David}} \yr{1983}  \at{Anisotropy in mhd turbulence due to a mean magnetic
  field}.  \jt{Journal of Plasma Physics}  \bvol{29}~(3),  \pg{525--547}.

\bibitem[{Spence}(2019)]{SpenceEA19}
{\sc \au{{Spence}, H.~E.}} \yr{2019} {HelioSwarm: Unlocking the Multiscale
  Mysteries of Weakly-Collisional Magnetized Plasma Turbulence and Ion
  Heating}.  \bt{In {\em AGU Fall Meeting Abstracts\/}}, ,  \vol{vol. 2019},
  \pg{pp. SH11B--04}.

\bibitem[Spyksma {\em et~al.\/}(2012)Spyksma, Magcalas \&
  Campbell]{spyksma2012quantifying}
{\sc \au{Spyksma, Kyle}, \au{Magcalas, Moriah} \& \au{Campbell, Natalie}}
  \yr{2012}  \at{Quantifying effects of hyperviscosity on isotropic
  turbulence}.  \jt{Physics of Fluids}  \bvol{24}~(12),  \pg{125102}.

\bibitem[Stawarz {\em et~al.\/}(2009)Stawarz, Smith, Vasquez, Forman \&
  MacBride]{Stawarz09}
{\sc \au{Stawarz, Joshua~E.}, \au{Smith, Charles~W.}, \au{Vasquez, Bernard~J.},
  \au{Forman, Miriam~A.} \& \au{MacBride, Benjamin~T.}} \yr{2009}  \at{The
  turbulent cascade and proton heating in the solar wind at 1 {AU}}.
  \jt{Astrophys.~J.}  \bvol{697},  \pg{1119--1127}.

\bibitem[Stawarz {\em et~al.\/}(2010)Stawarz, Smith, Vasquez, Forman \&
  Mac{Bride}]{StawarzEA10}
{\sc \au{Stawarz, J.~E.}, \au{Smith, C.~W.}, \au{Vasquez, B.~J.}, \au{Forman,
  M.~A.} \& \au{Mac{Bride}, B.~T.}} \yr{2010}  \at{The turbulent cascade for
  high cross-helicity states at 1\,au}.  \jt{Astrophys.\ J.}  \bvol{713},
  \pg{920--934}.

\bibitem[Stawarz {\em et~al.\/}(2011)Stawarz, Vasquez, Smith, Forman \&
  Klewicki]{Stawarz2011third}
{\sc \au{Stawarz, Joshua~E}, \au{Vasquez, Bernard~J}, \au{Smith, Charles~W},
  \au{Forman, Miriam~A} \& \au{Klewicki, Joseph}} \yr{2011}  \at{Third moments
  and the role of anisotropy from velocity shear in the solar wind}.  \jt{The
  Astrophysical Journal}  \bvol{736}~(1),  \pg{44}.

\bibitem[Taylor(1938)]{Taylor38}
{\sc \au{Taylor, G.~I.}} \yr{1938}  \at{The spectrum of turbulence}.
  \jt{Proc.\ Roy.\ Soc.\ Lond.\ A}  \bvol{164},  \pg{476--490}.

\bibitem[Taylor {\em et~al.\/}(2003)Taylor, Kurien \& Eyink]{TaylorEA03}
{\sc \au{Taylor, M.~A.}, \au{Kurien, S.} \& \au{Eyink, G.~L.}} \yr{2003}
  \at{Recovering isotropic statistics in turbulence simulations: {The}
  {Kolmogorov} 4/5th law}.  \jt{Phys.\ Rev.\ E}  \bvol{68},  \pg{026310}.

\bibitem[Tu \& Marsch(1995)]{TuMarsch95}
{\sc \au{Tu, C.-{Y}.} \& \au{Marsch, E.}} \yr{1995}  \at{\mhd structures, waves
  and turbulence in the solar wind}.  \jt{Space Sci.\ Rev.}  \bvol{73},
  \pg{1--210}.

\bibitem[Verdini {\em et~al.\/}(2015)Verdini, Grappin, Hellinger, Landi \&
  M{\"u}ller]{verdini2015anisotropy}
{\sc \au{Verdini, Andrea}, \au{Grappin, Roland}, \au{Hellinger, Petr},
  \au{Landi, Simone} \& \au{M{\"u}ller, Wolf~Christian}} \yr{2015}
  \at{Anisotropy of third-order structure functions in mhd turbulence}.
  \jt{The Astrophysical Journal}  \bvol{804}~(2),  \pg{119}.

\bibitem[{Verdini} {\em et~al.\/}(2015){Verdini}, {Grappin}, {Hellinger},
  {Landi} \& {M{\"u}ller}]{VerdiniEA15}
{\sc \au{{Verdini}, Andrea}, \au{{Grappin}, Roland}, \au{{Hellinger}, Petr},
  \au{{Landi}, Simone} \& \au{{M{\"u}ller}, Wolf~Christian}} \yr{2015}
  \at{{Anisotropy of Third-order Structure Functions in MHD Turbulence}}.
  \jt{Astrophys. J.}  \bvol{804}~(2),  \pg{119},  \arxiv{arXiv: 1502.04705}.

\bibitem[Wan {\em et~al.\/}(2009)Wan, Servidio, Oughton \&
  Matthaeus]{Wan2009third}
{\sc \au{Wan, Minping}, \au{Servidio, Sergio}, \au{Oughton, Sean} \&
  \au{Matthaeus, William~H}} \yr{2009}  \at{The third-order law for increments
  in magnetohydrodynamic turbulence with constant shear}.  \jt{Physics of
  plasmas}  \bvol{16}~(9),  \pg{090703}.

\bibitem[Wan {\em et~al.\/}(2010)Wan, Servidio, Oughton \&
  Matthaeus]{Wan2010third}
{\sc \au{Wan, Minping}, \au{Servidio, Sergio}, \au{Oughton, Sean} \&
  \au{Matthaeus, William~H}} \yr{2010}  \at{The third-order law for
  magnetohydrodynamic turbulence with shear: Numerical investigation}.
  \jt{Physics of Plasmas}  \bvol{17}~(5),  \pg{052307}.

\bibitem[Wang {\em et~al.\/}(2022)Wang, Chhiber, Adhikari, Yang, Bandyopadhyay,
  Shay, Oughton, Matthaeus \& Cuesta]{wang2022strategies}
{\sc \au{Wang, Yanwen}, \au{Chhiber, Rohit}, \au{Adhikari, Subash}, \au{Yang,
  Yan}, \au{Bandyopadhyay, Riddhi}, \au{Shay, Michael~A}, \au{Oughton, Sean},
  \au{Matthaeus, William~H} \& \au{Cuesta, Manuel~E}} \yr{2022}  \at{Strategies
  for determining the cascade rate in mhd turbulence: isotropy, anisotropy, and
  spacecraft sampling}.  \jt{The Astrophysical Journal}  \bvol{937}~(2),
  \pg{76}.

\bibitem[Yang {\em et~al.\/}(2021)Yang, Linkmann, Biferale \&
  Wan]{yang2021effects}
{\sc \au{Yang, Yan}, \au{Linkmann, Moritz}, \au{Biferale, Luca} \& \au{Wan,
  Minping}} \yr{2021}  \at{Effects of forcing mechanisms on the multiscale
  properties of magnetohydrodynamics}.  \jt{The Astrophysical Journal}
  \bvol{909}~(2),  \pg{175}.

\bibitem[Yang {\em et~al.\/}(2017)Yang, Matthaeus, Shi, Wan \&
  Chen]{YangEA-POF-17}
{\sc \au{Yang, Y.}, \au{Matthaeus, W.~H.}, \au{Shi, Y.}, \au{Wan, M.} \&
  \au{Chen, S.}} \yr{2017}  \at{Compressibility effect on coherent structures,
  energy transfer and scaling in magnetohydrodynamic turbulence}.
  \jt{Phys.~Fluid}  \bvol{29},  \pg{035105}.

\bibitem[Yokoyama \& Takaoka(2021)]{yokoyama2021energy}
{\sc \au{Yokoyama, Naoto} \& \au{Takaoka, Masanori}} \yr{2021}  \at{Energy-flux
  vector in anisotropic turbulence: application to rotating turbulence}.
  \jt{Journal of Fluid Mechanics}  \bvol{908}.

\end{thebibliography}

\end{document}